\documentclass{JFM-FLM_Au}

\usepackage{amsmath,amssymb,bm,mathtools}
\usepackage{natbib}
\usepackage{xcolor}
\usepackage[normalem]{ulem}

\providecommand{\st}[1]{\sout{#1}}
\newcommand{\newold}[2]{{\color{blue}{}#1}\if\relax\detokenize{#2}\relax\else\space{{\color{teal}\st{#2}}}\fi}

\newcommand{\defeq}{\overset{\text{def}}{=}}

\newcommand{\Ro}{\mbox{\textit{Ro}}}
\newcommand{\Fr}{\mbox{\textit{Fr}}}
\newcommand{\Eu}{\mbox{\textit{Eu}}}
\newcommand{\Bu}{\mbox{\textit{Bu}}}

\lefttitle{L. R. Seitz and B. A. Wingate}
\righttitle{Journal of Fluid Mechanics}

\title{A unifying view of averaging for wave--mean flow decomposition}
\author{L. R. Seitz\aff{1} and Beth A. Wingate\aff{2}}
\affiliation{\aff{1} Division of Applied Mathematics, Brown University, Providence, RI 02906, USA
\aff{2} Department of Mathematics, University of Exeter, Harrison Building, Exeter EX4 4PY, UK}
\corresau{L. R. Seitz, \texttt{lulabel\_seitz@brown.edu}}

\renewcommand{\pagelimitfooter}{}

\makeatletter
\let\oddabsfooterflag\@oddfoot
\let\evenabsfooterflag\@evenfoot
\makeatother

\usepackage{arxiv-layout}

\begin{document}
\maketitle

\begin{abstract}
Geophysical flows are routinely decomposed into mean flow and waves, and the procedure used determines how the motion is divided between the two. Ideally, the wave and mean motions of interest evolve on distinct timescales, in which case they can be separated by taking an infinite-time average over all wave periods. Individual waves average away, while exactly resonant interactions contribute to the averaged dynamics because they are steady on the fast timescale. We generalize this averaging to finite windows, yielding a family of decompositions that additionally retain near-resonant interactions whose phases do not fully cancel over the time window. Such averaging necessarily invokes a choice of reference frame; the same flow averaged in two frames may yield mean flows differing by a drift. The averaging formulation developed here enables us to express Stokes drift through resonant and near-resonant interactions, confirming the intuition that it is a slow quantity generated by fast waves and revealing a coherent near-resonant contribution generally present in Stokes drift calculated over finite time from broadband wave fields. We use the fast-wave evolution underlying the exact average to define both the wave reference frame, in which the fast phase is steady, and a wave-frame average. We cast familiar Eulerian and Lagrangian averages and this wave-frame average in a common form to further illustrate how the chosen frame determines which motions enter the mean. This view of averaging provides a common basis for comparing decompositions and choosing among them according to the physical processes one seeks to isolate.
\end{abstract}

\section{Introduction}\label{sec:introduction}

Separating the slow, balanced component of a flow from its fast wave oscillations is among the most fundamental operations in geophysical fluid dynamics \citep[e.g.][]{charney1948scale,gill1982atmosphere,warnEtAl1995Rossby,salmon1998lecturesGFD,julienKnobloch2007,vanneste2013balance,buhler2014wavesMeanFlows,vallis2017aofd2e} --- yet the question of how to perform this separation is deceptively simple, as different averaging procedures can produce different mean flows, different wave fields, and different accounts of how the two interact (\citealp{andrewsMcIntyre1976EP,andrewsmcintyre1978a}; \citealp{vanneste2022stokes}). There is no shortage of methods for wave--mean flow decomposition, from foundational formulations \citep{eliassenpalm1961,brethertonGarrett1968wavetrains,whitham1974linearNonlinearWaves,andrewsMcIntyre1976EP,andrewsmcintyre1978a,embidmajda1996} to more recent wave-averaged theories \citep{wagneryoung2015} and diagnostic and computational approaches for separating wave and mean components \citep{earlyLelongSundermeyer2021waveVortex,kafiabad2022,bakerkafiabadmaitlanddaviesvanneste2025}. These methods are each physically or practically well-motivated, but it is not always clear whether a given procedure achieves the intended separation between the waves and the mean flow for the application at hand. When the separation is incomplete, it is natural to ask what remains and what dynamical effect it has.
\begin{figure}[t]
  \centering
  \includegraphics[width=\linewidth]{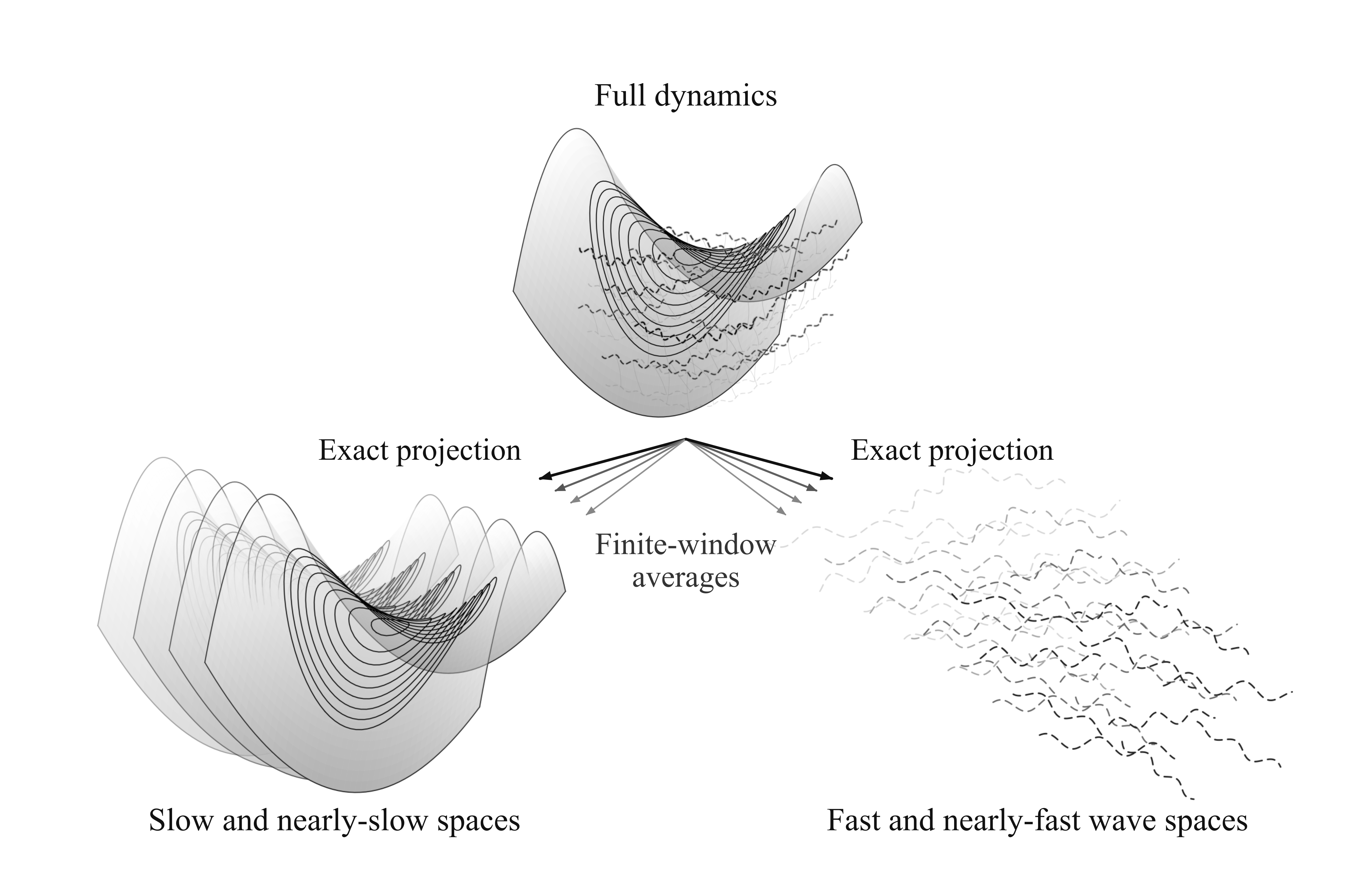}
  \caption{Schematic illustration of the many ways to separate a flow into a slow, balanced mean component and a fast, oscillatory wave component. The infinite-time average, or exact projection, selects one slow space and one complementary fast wave space; finite-window averages relax that exact split, producing a family of nearby slow and nearly-slow parts together with a corresponding family of fast and nearly-fast parts. Stokes drift, though generated by the waves, is a mean quantity that lives in a slow space (here, slow is meant in the sense of zero fast-time frequency, and this space need not coincide with the original balanced slow space). When calculating Stokes drift, replacing the exact average, which can be viewed as an average over infinite time, with one taken instead over a finite time window can add an extra, nearly-slow piece to the drift---one the infinite-time average, having discarded that near-resonant content, does not produce.}
  \label{fig:schematic}
\end{figure} 

Several established families of wave--mean decomposition implement this separation in different ways. Perhaps the oldest and most elementary is the Reynolds decomposition (\citealp{reynolds1895}; see \citealp{tennekesLumley1972firstCourse} and \citealp{pope2000turbulentFlows} for standard textbook treatments), which separates a flow into its time, space, or ensemble mean and the fluctuations about that mean. However, partitioning in this fashion need not isolate wave motions from other fluctuations, including turbulence and balanced motion carrying potential vorticity (\citealp{hasselmann1971massMomentumTransfer}; \citealp{earlyLelongSundermeyer2021waveVortex}). The transformed Eulerian mean, developed from Eliassen--Palm theory \citep{eliassenpalm1961,andrewsMcIntyre1976EP}, addresses some of these limitations while remaining within an Eulerian reference frame by reformulating the mean equations so that the residual circulation better represents wave-driven transport \citep[see, for example,][]{edmonHoskinsMcIntyre1980EPCrossSections,andrewsHoltonLeovy1987middleAtmDyn}. Ray and wave-action theory (\citealp{bretherton1966groupsShear,brethertonGarrett1968wavetrains,whitham1974linearNonlinearWaves}; see \citealp{grimshaw1984waveActionReview} for a review) averages over the fast phase of a slowly modulated wavetrain and describes its slow evolution through equations for the local frequency, wavenumber, and amplitude, the amplitude set by conservation of wave action; for internal gravity waves, the same approach also gives the mean motion the waves induce in the surrounding flow \citep{bretherton1969meanMotion}. Phase averaging in Hamilton's principle, together with asymptotic expansion, gives coupled equations for internal waves and the mean flow in a rotating stratified fluid \citep{gjaja1996}. Lagrangian-mean approaches, notably generalized Lagrangian mean (GLM) theory (\citealp{andrewsmcintyre1978a,andrewsmcintyre1978b}; see \citealp{mcintyre1980glmIntro} for an introduction, \citealp{holm2002} and \citealp{salmon2013alternative} for variational formulations, \citealp{gilbertVanneste2018geometric} for a geometric generalisation and \citealp{gilbertVanneste2025geometric} for a review) and its wave-averaged descendants \citep{wagneryoung2015,xieVanneste2015NIWGLM,salmon2016variational,thomasbuhlershafersmith2018,kafiabadvannesteyoung2021}, define the mean flow in terms of fluid-parcel trajectories, naturally preserving material conservation laws. Recent numerical methods compute Lagrangian means directly from gridded flow fields without explicit particle tracking \citep{kafiabad2022,kafiabadvanneste2023}, with subsequent extensions to general convolutional filters \citep{bakerkafiabadmaitlanddaviesvanneste2025} and efficient exponential filters \citep{minzbakerkafiabadvanneste2025}. When available, linear normal modes define an exact decomposition of the instantaneous state into its nonzero-frequency inertia--gravity-wave components and its zero-frequency geostrophic (vortical) component (\citealp{leith1980,bartello1995geostrophicAdjustment}; \citealp{earlyLelongSundermeyer2021waveVortex}; applications include \citealp{smithWaleffe2002slowLargeScales,waiteBartello2006geostrophicStratified}). Across these families, many methods can be expressed as time averages, often over a window chosen to span a particular wave period or set of periods, which motivates our focus on time averaging throughout this paper.

\begin{figure}[t]
  \centering
  \includegraphics[width=\linewidth]{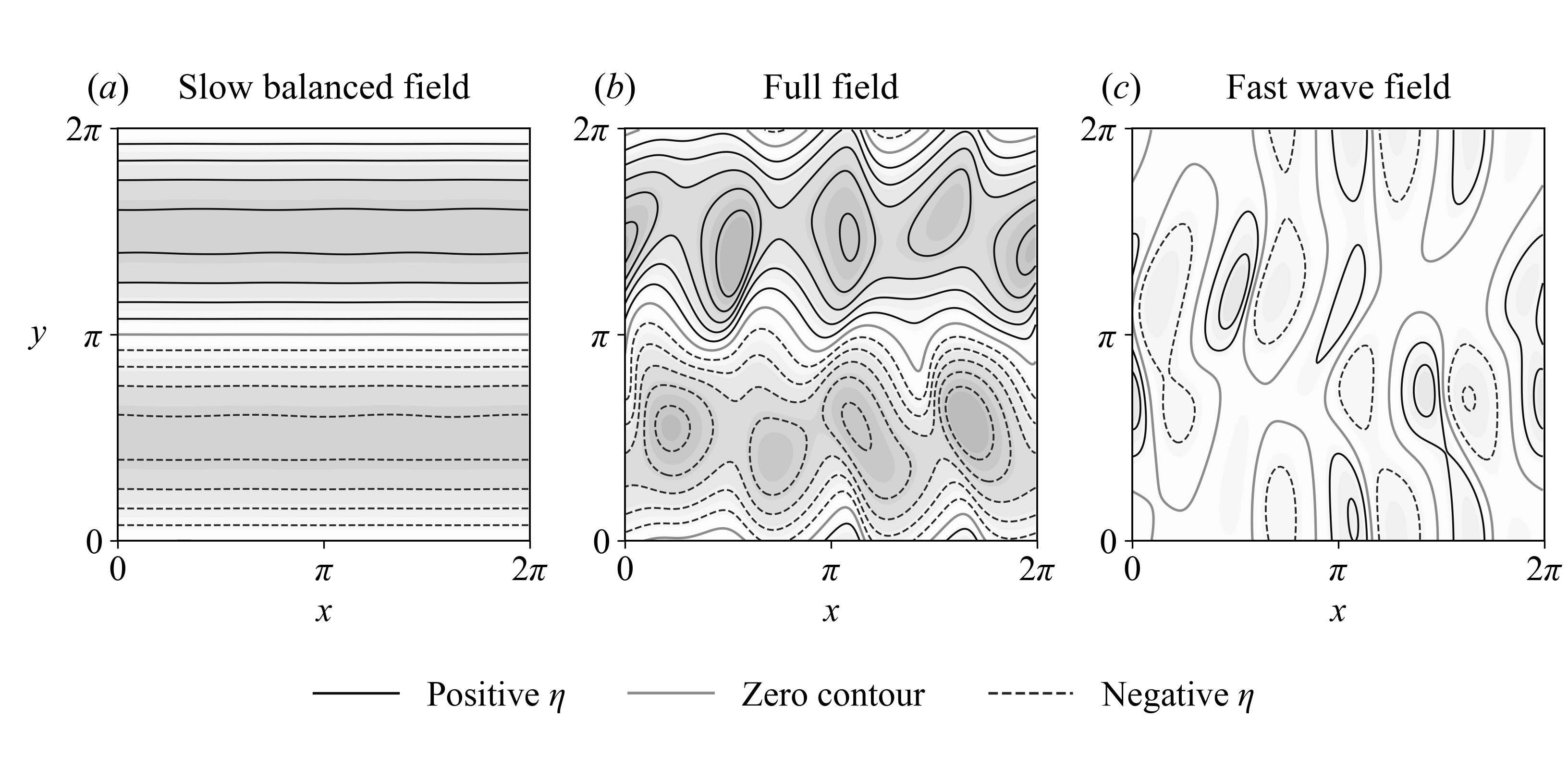}
  \caption{A concrete example of the exact slow--fast split in figure~\ref{fig:schematic} is provided by the free-surface height $\eta$ from a simulation of the two-dimensional rotating shallow water equations (detailed in \S\ref{sec:numerical-simulations}), with the illustration drawing on \citet{peddleHautWingate2019parareal}. ($a$)~The slow balanced field, obtained by the exact projection onto the geostrophic subspace. ($b$)~The full field, containing both balanced flow and inertia--gravity waves. ($c$)~The fast wave field, obtained by subtracting ($a$) from ($b$).}
  \label{fig:peddle-decomposition}
\end{figure}

One way to achieve an exact wave--mean decomposition of the state is to average over all wave periods, which amounts to taking an infinite-time average. This is the classical method of averaging \citep[see, among others,][]{mitropolsky1967,nayfeh1993introduction,sandersVerhulstMurdock2007averaging}. In this limit, oscillatory contributions vanish by the Riemann--Lebesgue lemma, leaving only the component invariant under the fast evolution. In the averaged dynamics, this amounts to retaining only the exactly resonant nonlinear interactions \citep[see, for example,][]{bretherton1964discreteOscillations,chenChenEyinkHolm2005resonant,wingateEtAl2011lowRossby}. We make explicit that the infinite-time state average over the fast linear evolution and the spectral projection onto the fast operator's zero-frequency subspace are equivalent (Appendix~\ref{app:spectral-projection}). The resulting exact slow--fast split is shown schematically in figure~\ref{fig:schematic} and with rotating shallow water fields in figure~\ref{fig:peddle-decomposition}. Within the theory of fast singular limits \citep{embidmajda1996,babin1997asymptotic,embidmajda1998,babin2000primitive}, the projection that produces this separation identifies the balanced state governed by the reduced equations. The scaling regime and its associated fast dynamics determine which balanced system is obtained. When, as is often the case, the scaling regime is chosen to reflect a particular balance of forces, the reduction map identifies a mean whose persistence under the fast dynamics follows from that balance. For instance, the classical quasi-geostrophic (QG) equations \citep{charney1948scale} arise under the joint limit of small Rossby and Froude numbers (\citealp{embidmajda1996,embidmajda1998}; see \citealp{julienKnoblochMilliffWerne2006} for generalized anisotropic regimes).

In this work, we generalize this infinite-time average to finite windows, producing a family of wave--mean partitions (figure~\ref{fig:schematic}) whose retained wave content depends on how completely the wave oscillations cancel over the averaging window. Replacing exact projection with finite-time averaging has a computational counterpart in optimal balance, where a finite-time average has been used to approximate the exact geostrophic projection applied at the linear stage \citep{masurOliver2020}, with retained wave content treated as projection error \citep{rosenauChoukseyEden2026}. Here, however, when averaging nonlinear terms involving waves, the near-resonant content retained by a finite-window average need not merely constitute error, because it can preserve physical effects removed by infinite-time averaging.

The dynamical importance of near-resonant interactions is well established, from analysis of Rossby wave packets \citep{newell1969rossbyWavePacket} to later studies in rotating flows \citep{smithLee2005nearResonances,bourouibaBartello2007intermediateRossby,leeSmith2007zonalFlowsNearResonant,clarkDiLeoniMininni2016nearResonant} and broader wave-turbulence theory \citep{newellRumpf2011waveTurbulence}. Near-resonant interactions can transfer energy and generate slow modes without satisfying the exact resonance condition, and retaining such interactions can improve the accuracy of reduced models. In rotating and beta-plane flows, reduced models restricted to a near-resonant band of nonlinear triads capture large-scale transfer and zonal-flow formation that stricter exact-resonant descriptions cannot explain \citep{smithLee2005nearResonances,leeSmith2007zonalFlowsNearResonant}. The same finite-time resonance-broadening appears for surface gravity waves, where a nonresonant extension of four-wave kinetic theory compares favourably with direct simulations \citep{janssen2003freakWaves} and exact-resonance-only calculations can be physically meaningless unless approximately resonant interactions are retained \citep{annenkovShrira2006nonresonantRandomWave}. Wave--mean decompositions based on averaging over finite time windows, as is common in practice, may therefore contain dynamical content that the classical average would not categorize as part of the mean flow.

However, averaging an oscillatory field depends not only on the length of time over which the average is taken, but also on the reference frame in which the average is computed. A simple example of this reference-frame dependence is provided by near-inertial motion, whose horizontal velocity vector rotates rapidly at the Coriolis frequency even when the amplitude and phase of the motion evolve slowly. In the original reference frame, defined by the usual spatial coordinates, the average extracts the slowly varying background current, whereas in a reference frame rotating with the inertial oscillation, the same average extracts the slowly varying near-inertial component. To show how the same averaging window acts in the two frames, consider a horizontal-velocity time series recorded at a fixed spatial location. The horizontal velocity $(u(t),v(t))$ may be represented by a single complex variable $\mathcal U(t)=u(t)+iv(t)$, which we take to have the form
\begin{equation}
  \mathcal U(t)=U(T)+A(T)e^{-ift}.
\end{equation}
Here $f$ is the Coriolis frequency and $T=\varepsilon t$, with $0<\varepsilon\ll 1$, is a slow time on which the background current $U(T)$ and complex near-inertial amplitude $A(T)$ vary.

A direct average over one inertial period in the original frame gives
\begin{equation}\label{eq:intro-niw-direct-average}
\begin{aligned}[b]
  \overline{\mathcal U}^{\mathrm{direct}}(t) &\defeq \frac{|f|}{2\pi}\int_t^{t+2\pi/|f|}\mathcal U(s)\,\mathrm{d}s \\
  &= U(T)+A(T)\left(\frac{|f|}{2\pi}\int_t^{t+2\pi/|f|}e^{-ifs}\,\mathrm{d}s\right)+O(\varepsilon) \\
  &= U(T)+O(\varepsilon).
\end{aligned}
\end{equation}
Thus the direct average extracts the slow background current but discards the $O(1)$ near-inertial amplitude $A$.

We can instead express the velocity in a frame rotating with the near-inertial oscillation by defining
\begin{equation}\label{eq:intro-niw-mapped-variable}
  \mathcal V(t)\defeq e^{ift}\mathcal U(t).
\end{equation}
In this frame, the near-inertial component is slow, so averaging over one inertial period gives
\begin{equation}\label{eq:intro-niw-mapped-average}
\begin{aligned}[b]
  \overline{\mathcal U}^{\mathrm{mapped}}(t) &\defeq \frac{|f|}{2\pi}\int_t^{t+2\pi/|f|}\mathcal V(s)\,\mathrm{d}s \\
  &= U(T)\left(\frac{|f|}{2\pi}\int_t^{t+2\pi/|f|}e^{ifs}\,\mathrm{d}s\right)+A(T)+O(\varepsilon) \\
  &= A(T)+O(\varepsilon).
\end{aligned}
\end{equation}
Hence, after a change of frame, the same averaging operation extracts the near-inertial amplitude $A$ rather than the background current $U$. An observer seeking to isolate the near-inertial signal from a velocity record therefore averages in the rotating frame, as is done both in near-inertial wave theory \citep{youngBenJelloul1997} and in observational practice \citep{goughEtAl2016,cyriacEtAl2022}. Transforming away known fast linear motion before averaging is a standard construction in classical averaging \citep[sections~1.6--1.7]{sandersVerhulstMurdock2007averaging}.

In general, what is identified as fast wave content is a matter of reference frame, so any averaging procedure that removes fast oscillations necessarily invokes a choice of frame. An Eulerian average is taken at a fixed point and a Lagrangian average follows a fluid parcel; we call the frame that follows the fast wave oscillation through its evolving phase the \emph{wave reference frame}. When viewed in this frame, the wave phase no longer varies on the fast timescale. In the near-inertial example, the mapping \eqref{eq:intro-niw-mapped-variable} implements precisely this change of frame, under which the same averaging operation identifies a different quantity as slow. Here the wave is the near-inertial velocity component $A(T)e^{-ift}$, and the corresponding frame uses horizontal axes rotating with its fast inertial phase, leaving the slow modulation of $A(T)$ unchanged.

\begin{figure}[t]
  \centering
  \includegraphics[width=0.7\linewidth]{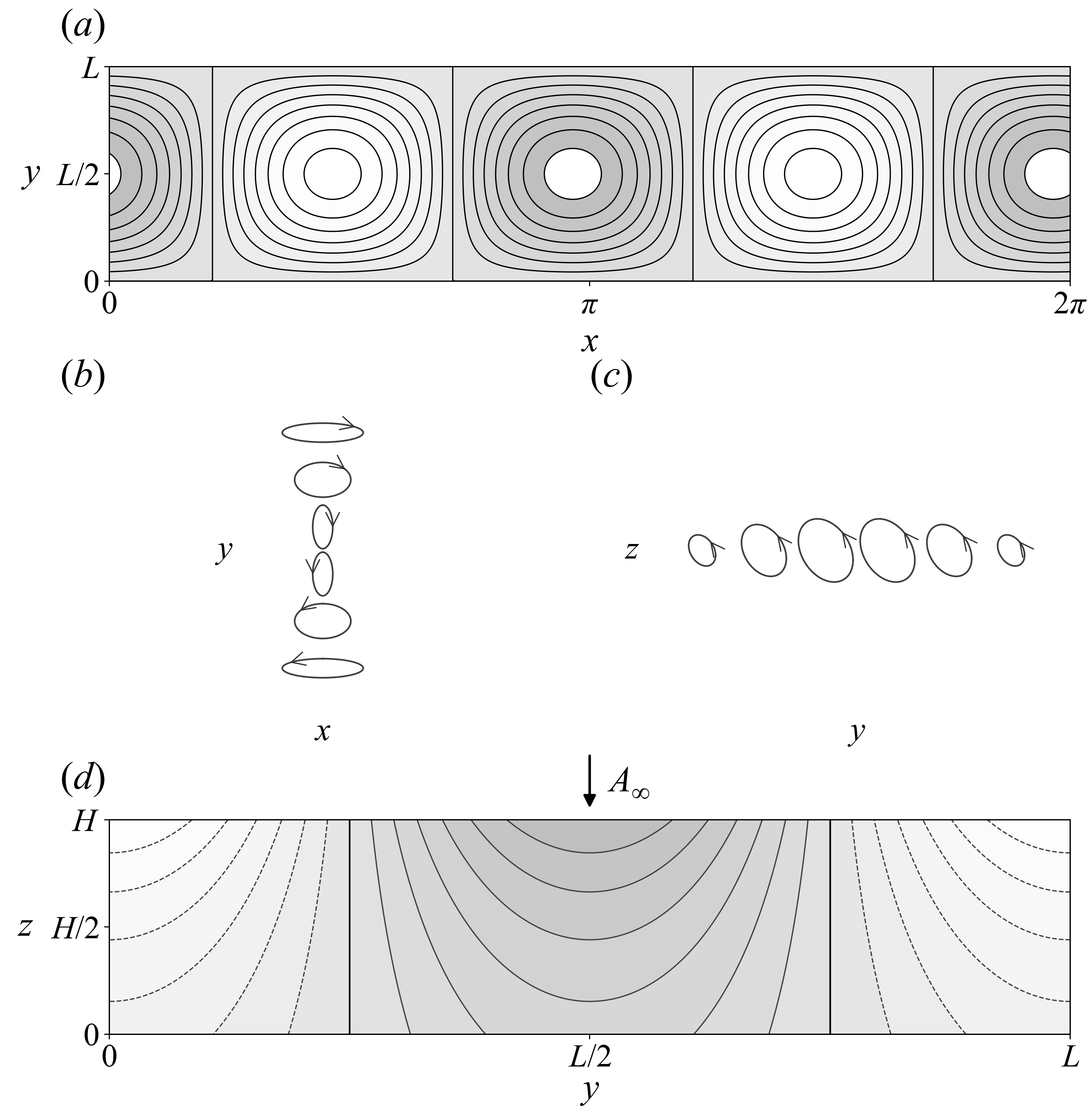}
  \caption{The figure follows the classical stationary Rossby-wave example from its Eulerian wave field to the resulting parcel orbits and Stokes drift. Panel~(a) shows the Eulerian wave field through the streamfunction $\psi'$ in the $x$--$y$ plane. Panels~(b) and (c) show the first-order parcel orbits in the $x$--$y$ and $y$--$z$ planes, calculated following \citet{andrewsHoltonLeovy1987middleAtmDyn}, figure~4.13. For visual clarity, the parcel displacements are enlarged, with the zonal displacement in panel~(b) drawn at four times the meridional scale. Panel~(d) shows the zonal component of the classical Stokes drift derived from the same wave in the $y$--$z$ plane, extending the textbook example.}
  \label{fig:andrews-single-wave}
\end{figure}

Introducing the wave reference frame naturally raises the question of drift. When a reference frame is tied to fast oscillatory motion, the coordinate transformation between frames can introduce nonlinear dependence on the wave field, and averaging such a nonlinear function of oscillatory quantities generically produces a nonzero residual --- the drift. This mechanism is ubiquitous across physics and mathematics (\citealp[e.g.,][]{lighthill1978acousticStreaming,sandersVerhulstMurdock2007averaging,caryBrizard2009guidingCentre,eamesMcIntyre1999stokesDarwin,kapitza1951vibrationalStabilisation,craikLeibovich1976Langmuir}; \citealp{holm1996}). In wave--mean interaction theory, the most familiar example is Stokes drift, the difference between the Lagrangian- and Eulerian-mean velocities, first described by \citet{stokes1847oscillatoryWaves}. Because it is a quadratic quantity built from the wave field \citep[see, e.g.,][]{longuetHiggins1953massTransport,vandenBremerBreivik2018stokesDrift}, it is directly sensitive to which nonlinear interactions contribute to the averaged field. However, Stokes drift is only one of many physical drift quantities associated with a change of reference frame in oscillatory flows; more broadly, acoustic streaming \citep{lighthill1978acousticStreaming}, Darwin drift \citep{eamesMcIntyre1999stokesDarwin}, and the mean water-level changes and currents driven by gradients in radiation stress (\citeauthor{longuetHigginsStewart1964radiationStress} \citeyear{longuethigginsstewart1962}, \citeyear{longuetHigginsStewart1964radiationStress}) are equally familiar examples of related mean effects induced by waves and moving bodies. The appearance of additional terms in the averaged equations for slow motion is also a generic feature of ordinary and partial differential equations with fast oscillations; the Kapitza pendulum exhibits an effective slow potential from fast pivot oscillation \citep{kapitza1951vibrationalStabilisation}, elimination of fast velocity in Langevin dynamics can produce an additional drift \citep{hottovyVolpeWehr2012noiseInducedDrift}, and the Bogoliubov--Krylov method organizes the whole phenomenon systematically for ordinary differential equations \citep{sandersVerhulstMurdock2007averaging}. A normal-form treatment of slow--fast systems links slow drift, including Stokes drift, to quadratic interactions between fast waves (\citealp{roberts2008normal}; \citealp[\S2.5]{roberts2022backwards}).

Stokes drift, which arises from correlations between wave-induced displacement and oscillatory velocity gradients, can be written for surface gravity waves in terms of interactions between wave harmonics. Under an infinite-time average, only equal-frequency pairs remain (\citealp{nouguierchapronguerin2015}); unequal-frequency pairs can instead produce low-frequency Stokes-drift fluctuations \citep{herbersjanssen2016}. In this paper, we express Stokes drift as a sum of contributions from pairs of fast wave modes, revealing that the pairs retained by exact averaging satisfy the frequency condition for a fast--fast--slow resonance, a connection made explicit through the averaging formulation developed here. We further show that a finite-window average retains pairs with nearby but unequal frequencies, adding a distinct, window-dependent contribution to the Stokes drift that we call the \emph{nearly-slow drift}.

To illustrate how the nearly-slow drift arises, we begin with a classical example of Lagrangian averaging: a single stationary Rossby wave whose vertical propagation produces parcel displacements and a nonzero Stokes drift \citep{andrewsHoltonLeovy1987middleAtmDyn} (figure~\ref{fig:andrews-single-wave}). For the full-period window used here, the finite-window and infinite-time drifts associated with this single wave agree. Adding a second wave at a nearby frequency introduces an interaction that varies at the small frequency difference and produces a coherent nearly-slow drift. Importantly, this drift, which is generated by near-resonant interactions retained by the finite-window but not the infinite-time average, causes the predicted parcel orbits to separate, whereas they coincide in the single-wave case shown in figure~\ref{fig:andrews-single-wave}(b). Unlike classical infinite-time averaging, finite-window averaging generally does not define a projection, so the resulting decomposition is not canonical: each window length defines its own mean flow and drift. The choice of Lagrangian mean map is itself a source of non-uniqueness for Stokes drift \citep{vanneste2022stokes}; here, the finite-window average adds a distinct dependence on what the window retains. This is one manifestation of a broader non-uniqueness explored in \S\ref{sec:exact-and-finite} and \S\ref{sec:numerical-simulations}.

The remainder of this paper is organized as follows. In \S\ref{sec:governing-fast-structure}, we introduce the rotating Boussinesq equations and their fast-time structure, together with the wave reference frame generated by the fast linear dynamics and the corresponding exact projection. The finite-window formulation developed in \S\ref{sec:exact-and-finite} shows how incomplete phase cancellation retains window-dependent near-resonant content in the identified mean. In \S\ref{sec:drift}, we derive Stokes drift under exact and finite-window averaging, showing that near-resonant interactions retained by the finite window add a nearly-slow transport contribution. We illustrate in \S\ref{sec:numerical-simulations} how the averaging window and reference frame determine the resulting wave--mean partition, using three numerical examples. The first compares the exact and finite-window drifts produced after adding a nearby second frequency to the stationary Rossby-wave example in figure~\ref{fig:andrews-single-wave} and isolates their difference as a coherent nearly-slow contribution whose weight varies with the averaging window. The second compares direct finite-window averaging of a rotating shallow water flow with the same averaging after the flow is mapped into the wave reference frame, extending the reference-frame dependence of near-inertial motion into a fully nonlinear setting. The third expresses Eulerian averaging, Lagrangian averaging, and infinite-time averaging over the fast-wave evolution in directly comparable forms and compares the wave--mean partitions they define. Finally, in \S\ref{sec:discussion}, we distinguish the exact separation of an instantaneous state into wave and mean, which the classical infinite-time average described here achieves, from the necessarily incomplete separation of nonlinear wave and mean dynamics, and place our results in a broader context drawn from several areas of physics.

\section{Governing equations and fast--slow structure}
\label{sec:governing-fast-structure}

To construct the wave reference frame and the exact average from the governing dynamics, we first write the rotating, stratified Boussinesq equations in fast--slow form. In particular, we consider the QG regime, where the fast linear motion consists of inertia--gravity waves.

\subsection{Governing equations}

The nondimensional rotating, stratified Boussinesq equations (specialized to vertical rotation) are given by
\begin{subequations}\label{eq:rotating-boussinesq-em98}
\begin{align}
\partial_t \bm{u} + \bm{u}\cdot\nabla \bm{u} + \Ro^{-1}\hat{\bm{e}}_z\times\bm{u} + \Eu\nabla\phi + \Gamma \rho\hat{\bm{e}}_z &= \Rey^{-1}\Delta\bm{u}, \label{eq:momentum-em98}\\
\partial_t \rho + \bm{u}\cdot\nabla\rho - \Gamma^{-1}\Fr^{-2}w &= \Rey^{-1}\Pran^{-1}\Delta\rho, \label{eq:density-em98}\\
\nabla\cdot\bm{u} &= 0, \label{eq:incompressible-em98}
\end{align}
\end{subequations}
where $\bm{u}=(u,v,w)$ is the velocity, $\rho$ is the density fluctuation, $\phi$ is pressure, and $\hat{\bm{e}}_z\defeq(0,0,1)^\top$. Using velocity scale $U$, length scale $L$, pressure scale $\Phi$, and density-fluctuation scale $\rho_0 B$, with $\rho_0$ the reference density, the Rossby, Froude, Reynolds, Prandtl, and Euler numbers arise naturally, together with the buoyancy-amplitude parameter $\Gamma$:
\begin{equation}\label{eq:boussinesq-dimensionless-parameters}
\Ro\defeq \frac{U}{fL}, \quad \Fr\defeq \frac{U}{NL}, \quad \Rey\defeq \frac{UL}{\nu}, \quad \Pran\defeq \frac{\nu}{\kappa}, \quad \Eu\defeq \frac{\Phi}{\rho_0 U^2}, \quad \Gamma\defeq \frac{B g L}{U^2}.
\end{equation}
Here, $f$ is the Coriolis frequency, $N$ is the Brunt--V\"ais\"al\"a frequency, $g$ is gravitational acceleration, $\nu$ is kinematic viscosity, and $\kappa$ is scalar diffusivity.

We now reorganize the system \eqref{eq:rotating-boussinesq-em98} into a simpler form that takes advantage of the timescale separation often underlying wave--mean decomposition. The term `mean flow' most simply evokes a sustained pattern of motion. Such a pattern is often maintained by an approximate balance of forces with waves arising as restoring motions when that balance is disturbed. The waves can then be most clearly distinguished from the mean when they evolve on a much shorter timescale, i.e. when the ratio $\varepsilon$ of the characteristic wave timescale to the mean-flow evolution timescale satisfies $\varepsilon\ll1$. With time $t$ measured in units of the advective timescale $L/U$, the terms governing these rapid wave motions enter at order $1/\varepsilon$. Concatenating $\bm{u}$ and $\rho$ into the state vector $\bm{w}\defeq(\bm{u},\rho)^\top$ allows us to write the system \eqref{eq:rotating-boussinesq-em98} as a single equation:
\begin{equation}\label{eq:abstract-system}
\frac{\partial \bm{w}}{\partial t} + \frac{1}{\varepsilon}\mathcal{L}_{\mathrm F} \bm{w} + \mathcal{L}_S \bm{w} + \mathcal{B}(\bm{w},\bm{w}) - \mathcal{D}(\bm{w}) = 0.
\end{equation}
In this representation, $\mathcal{L}_{\mathrm F}$ and $\mathcal{L}_S$ are the fast and slow linear operators, respectively, $\mathcal{B}$ is the bilinear operator representing nonlinear interactions, and $\mathcal{D}$ is the linear diffusion operator.

\subsection{Isolating the fast oscillations}

The scaling associated with the physical regime of interest determines which dynamics enter the leading fast operator. For example, a low-Froude scaling with~$\Ro=O(1)$ places stratification in the leading fast operator while rotation remains at order one.
In contrast, the QG specialization used here places both rotation and stratification in the leading fast balance by taking $\Fr=\varepsilon$, $\Ro=\varepsilon F^{-1}$, and $\Gamma=\Fr^{-1}=\varepsilon^{-1}$, with $F\defeq\Fr/\Ro=O(1)$, so that the Burger number $\Bu\defeq F^{-2}$ is also order one \citep[\S2]{embidmajda1998}. On a periodic domain, eliminating the pressure through the Leray projection gives the corresponding concrete operators in \eqref{eq:abstract-system}:
\begin{subequations}\label{eq:qg-operators-em98}
\begin{align}
\mathcal{L}_{\mathrm F} \bm{w} &\defeq \begin{pmatrix} F\hat{\bm{e}}_z\times\bm{u} + \rho\hat{\bm{e}}_z + \nabla\Delta^{-1}\bigl(F\hat{\bm{e}}_z\cdot(\nabla\times\bm{u}) - \partial_z\rho\bigr)\\ -w \end{pmatrix}, \label{eq:LF-qg-em98}\\
\mathcal{L}_S \bm{w} &\defeq 0, \label{eq:LS-qg-em98}\\
\mathcal{B}(\bm{w}_1,\bm{w}_2) &\defeq \begin{pmatrix} \bm{u}_1\cdot\nabla \bm{u}_2 - \nabla\Delta^{-1}\bigl(\nabla\cdot(\bm{u}_1\cdot\nabla \bm{u}_2)\bigr)\\ \bm{u}_1\cdot\nabla \rho_2 \end{pmatrix}. \label{eq:B-qg-em98}
\end{align}
\end{subequations}
Here, $\bm{w}_j\defeq(\bm{u}_j,\rho_j)^\top$, $j=1,2$, with $\bm{u}_1$ supplying the advecting velocity and $(\bm{u}_2,\rho_2)$ supplying the fields being advected.

The factor $1/\varepsilon$ multiplying $\mathcal{L}_{\mathrm F}$ in \eqref{eq:abstract-system} introduces the fast time $\tau \defeq t/\varepsilon$. With the multiple-scales expansion
\begin{equation}\label{eq:multiple-scales-expansion}
\bm{w} = \bm{w}^{(0)}(\bm{x},t,\tau) + \varepsilon\bm{w}^{(1)} + \cdots,
\end{equation}
the $O(\varepsilon^{-1})$ balance yields the leading-order equation
\begin{equation}\label{eq:leading-order}
\partial_\tau \bm{w}^{(0)} + \mathcal{L}_{\mathrm F} \bm{w}^{(0)} = 0,
\end{equation}
whose solution is
\begin{equation}\label{eq:leading-order-solution}
\bm{w}^{(0)}(\bm{x},t,\tau) = e^{-\tau\mathcal{L}_{\mathrm F}}\overline{\bm{w}}(\bm{x},t),
\end{equation}
for a slowly varying amplitude $\overline{\bm{w}}$. In other words, the leading-order state is oscillatory, but the amplitude varies only on the slow timescale. Note that, despite the abstract notation, $e^{-\tau\mathcal{L}_{\mathrm F}}$ in \eqref{eq:leading-order-solution} merely evolves the amplitude $\overline{\bm{w}}$ into the solution of the linear fast problem \eqref{eq:leading-order} over fast time $\tau$, and is often readily computable \citep[see, e.g.,][]{hautBabbMartinssonWingate2016timeEvolution,schreiberLoft2019parallelTimeIntegrator,yamazakiCotterWingate2023phaseAveraging}.

The amplitude $\overline{\bm{w}}$ is determined at the next order, where the correction $\bm{w}^{(1)}$ inherits a forcing from $\bm{w}^{(0)}$. Avoiding secular growth requires the fast-time average of that forcing to vanish, which yields a closed equation for the amplitude \citep[\S2]{embidmajda1998}:
\begin{equation}\label{eq:averaged-equation}
\frac{\partial \overline{\bm{w}}}{\partial t} + \lim_{T\to\infty}\frac{1}{T}\int_0^T e^{s\mathcal{L}_{\mathrm F}}\Bigl(\mathcal{L}_S e^{-s\mathcal{L}_{\mathrm F}}\overline{\bm{w}} + \mathcal{B}\bigl(e^{-s\mathcal{L}_{\mathrm F}}\overline{\bm{w}}, e^{-s\mathcal{L}_{\mathrm F}}\overline{\bm{w}}\bigr) - \mathcal{D}\bigl(e^{-s\mathcal{L}_{\mathrm F}}\overline{\bm{w}}\bigr)\Bigr)\mathrm{d}s = 0.
\end{equation}
Expanding $\overline{\bm{w}}$ in the oscillatory eigenmodes of $\mathcal{L}_{\mathrm F}$ shows which interactions enter the averaged equation: modes with frequencies $\omega_a$ and $\omega_b$ force, through $\mathcal{B}$, a mode of frequency $\omega_c$ with oscillating weight $e^{-is(\omega_a+\omega_b-\omega_c)}$, and the time average of this weight vanishes unless $\omega_a+\omega_b=\omega_c$ \citep[equations~(3.13)--(3.16)]{embidmajda1998}. These quadratic interactions therefore contribute only through exactly resonant triads.

According to \eqref{eq:leading-order-solution}, the leading-order state oscillates only through the factor $e^{-\tau\mathcal{L}_{\mathrm F}}$, so multiplying by $e^{\tau\mathcal{L}_{\mathrm F}}$ cancels the oscillation and returns the amplitude $\overline{\bm{w}}$. Applied to the full solution, this same transformation, as in \eqref{eq:intro-niw-mapped-variable}, expresses the state in coordinates that follow the evolving phase of the linear waves, defining the wave reference frame:
\begin{equation}\label{eq:wave-frame-variable}
\tilde{\bm{w}}(\bm{x},t) \defeq e^{(t/\varepsilon)\mathcal{L}_{\mathrm F}}\bm{w}(\bm{x},t).
\end{equation}
The change of variables removes the rapid phase generated by the leading fast problem without subtracting the wave content itself. A solution of the leading fast problem appears steady in $\tilde{\bm{w}}$, while the wave amplitudes remain present and can still evolve slowly.
Notice that the same change of frame already appears implicitly in \eqref{eq:averaged-equation}: it evaluates the slow linear, nonlinear, and diffusive terms on the reconstructed oscillating state, and then maps the result back into the wave reference frame before taking the fast-time average. For the full solution, substituting \eqref{eq:wave-frame-variable} into \eqref{eq:abstract-system} cancels the $1/\varepsilon$ term exactly (the `cancellation of oscillations' introduced by \citet{schochet1994} and extended to geophysical systems by \citet{embidmajda1996}).

The same fast linear operator that generates the wave reference frame also defines the corresponding infinite-time average, which, as we show here, projects onto the part of the state that carries no fast oscillation (that is, onto $\ker\mathcal{L}_{\mathrm F}$) and thereby gives the exact wave--mean decomposition of the instantaneous state for the chosen fast dynamics. When the fast linear oscillations conserve the energy they carry (mathematically, when $\mathcal{L}_{\mathrm F}$ is skew-adjoint in the energy inner product), the average is the orthogonal projection, which we denote by $P$, onto that fast-time-invariant part of the state (as proved in Appendix~\ref{app:spectral-projection}):
\begin{equation}\label{eq:exact-wave-filter}
P = \lim_{T\to\infty} \frac{1}{T}\int_0^T e^{-s\mathcal{L}_{\mathrm F}}\mathrm{d}s.
\end{equation}
In the multiple-scales derivation of \citet{embidmajda1996,embidmajda1998}, the infinite-time average arises from the requirement that the first-order correction remain free of secular growth. When the leading amplitude is subsequently restricted to $\ker\mathcal L_{\mathrm F}$, the resulting slow equation is written with the orthogonal projection $P$ acting on the remaining dynamical terms \citep{embidmajda1998}. One important consequence of this identification is that it allows us to generalize exact wave--mean separation to finite windows and characterize the additional content retained in the mean. A further consequence is to make explicit that, in regimes such as the QG limit, the classical average selects the component satisfying the underlying force balance.

The projection $P$ and the map into the wave reference frame in \eqref{eq:wave-frame-variable} are tied to the same linear wave motion, but use it in opposite directions. At the same fast-time value $\tau$, the map into the wave frame and the map appearing in $P$ are $e^{\tau\mathcal L_{\mathrm F}}$ and $e^{-\tau\mathcal L_{\mathrm F}}$, respectively, i.e. inverses of each other. The map into the wave frame applies the transformation $e^{\tau\mathcal L_{\mathrm F}}$ selected by the field's actual fast time $\tau=t/\varepsilon$, thereby undoing the phase accumulated under the leading fast evolution. The projection instead applies the family $e^{-\tau\mathcal L_{\mathrm F}}$ to one instantaneous field over all fast-wave phases and averages the resulting phase-shifted fields. The phase-dependent components cancel in this average, while the phase-invariant component remains. In the QG scaling of \eqref{eq:qg-operators-em98}, this invariant component is the balanced mean, with the horizontal velocity in geostrophic balance and the density in hydrostatic balance (equivalently, the waves carry no potential vorticity in this limit). The operator $P$ therefore projects the field onto that balanced mean. This also distinguishes the three quantities introduced in this section: $\tilde{\bm{w}}$, defined in \eqref{eq:wave-frame-variable}, is the full field rewritten in the wave reference frame; $\overline{\bm{w}}$, introduced in \eqref{eq:leading-order-solution} and governed by \eqref{eq:averaged-equation}, is the leading-order amplitude and still carries its resonant wave content; and $P\bm{w}$, with $P$ defined in \eqref{eq:exact-wave-filter}, is the balanced mean invariant under the leading fast wave dynamics.

\section{Averaging over finite time windows retains near-resonances}\label{sec:exact-and-finite}

The exact projection $P$ defined by \eqref{eq:exact-wave-filter} isolates the component of a field that is invariant under the fast linear motion. When the dispersion relation of the relevant waves is known, this projection can be readily constructed spectrally from the wave modes \citep{leith1980,bartello1995geostrophicAdjustment,embidmajda1996,babin1997asymptotic,embidmajda1998}. In complex geometries or with boundary conditions that do not admit simple normal modes, the wave modes may not be available analytically \citep[see, e.g.,][]{maasLam1997internalWaveAttractors,rieutordValdettaro1997inertialModes}, but a numerical approximation to the zero-frequency subspace may still be obtained from the fast operator. Although the exact projection may thus be numerically accessible, observational and numerical analyses typically use finite-window averages, raising the question of how a finite rather than infinite averaging interval alters the resulting wave--mean separation. We address this question by generalizing the averaging construction in \eqref{eq:exact-wave-filter} from its infinite-time limit to finite-time averages. This reflects a more general feature of finite-time dynamics: over a finite averaging interval, oscillations whose phase changes too little cannot be clearly distinguished from the mean.

In this section, we use the finite-time analogue to analyse how finite averaging time changes which contributions enter the mean. The distinct question of whether that same finite window is applied to the fields in place or after passage to the wave reference frame is deferred to \S\ref{sec:rsw-equations}.

\subsection{General form of time averages}

Once the field to be averaged is specified, any average depends on two choices: which values are included and how those values are weighted relative to one another. Hence, we may regard the finite-time generalization of the exact average in \eqref{eq:exact-wave-filter} as one instance of a more general class of time averages. For a field $f(t)$ evolving continuously in time, both choices may be encoded by a weighting function, or kernel, of the offset $s$ from an evaluation time $t_0$. To ensure that averaging preserves constants, we choose a kernel $K\in L^1(\mathbb{R})$ with unit integral. Writing $K_T(s)\defeq T^{-1}K(s/T)$ separates the two choices: $T$ sets the scale of the averaging window, while $K$ sets the relative weighting within it. At each offset, the kernel $K_T(s)$ gives the weight assigned to $f(t_0+s)$. The corresponding averaging operator is
\begin{equation}\label{eq:kernel-average}
(A_T f)(t_0) \defeq \int_{\mathbb{R}} K_T(s) f(t_0+s)\mathrm{d}s.
\end{equation}

For example, to take a simple time average over an interval of length $T$ centered on $t_0$, we assign every value inside the interval equal weight and every value outside it zero weight. These choices give the boxcar kernel $K\defeq\mathbf{1}_{[-1/2,1/2]}$, with $K_T(s)=T^{-1}\mathbf{1}_{[-T/2,T/2]}(s)$; substituting this kernel into \eqref{eq:kernel-average} gives
\begin{equation}
(A_T f)(t_0) = \frac{1}{T}\int_{-T/2}^{T/2} f(t_0+s)\mathrm{d}s. \label{eq:centered-uniform-average}
\end{equation}

The scaling in \S\ref{sec:governing-fast-structure} fixes the fast evolution, thereby determining both the wave reference frame and the associated infinite-time average \eqref{eq:exact-wave-filter}. To average over this fast motion for a finite duration, we apply \eqref{eq:kernel-average} to $f(t)\defeq e^{-t\mathcal L_{\mathrm F}}\bm w_0$, the fast orbit generated from an instantaneous state $\bm w_0$:
\begin{equation}\label{eq:finite-fast-orbit-average}
(A_T f)(0)=\int_{\mathbb R}K_T(s)e^{-s\mathcal L_{\mathrm F}}\bm w_0\mathrm{d}s.
\end{equation}
Choosing $K_T(s)=T^{-1}\mathbf1_{[0,T]}(s)$ recovers the uniform orbit average whose limit as $T\to\infty$ gives $P\bm w_0$, as in \eqref{eq:exact-wave-filter}. The product $K_T(s)e^{-s\mathcal L_{\mathrm F}}$ may itself be regarded as an operator-valued kernel, combining the scalar temporal weighting with the fast evolution \citep[e.g.][]{girardiweis2004}.

Holding the fast evolution fixed, we obtain a family of finite averages by varying the scalar temporal weighting $K_T$. In particular, the exact average removes every contribution that is not stationary relative to the selected fast oscillations, whereas a finite average can retain contributions that vary only slowly over the chosen window. Unlike $P$, $A_T$ is generally not a projection, so applying the same average again can further modify the partially retained components.

\subsection{Frequency selection by finite-time averaging}

We now quantify this distinction through the frequency response of the averaging operator \eqref{eq:kernel-average}. To begin with the simplest case, we apply $A_T$ to a function consisting of a single oscillatory component, $f(t)\defeq e^{-i\omega t}$, and evaluate its average at time $t_0$. In this case, \eqref{eq:kernel-average} gives
\begin{equation}\label{eq:spectral-multiplier}
(A_T f)(t_0) = e^{-i\omega t_0}\int_{\mathbb{R}}K_T(s)e^{-i\omega s}\mathrm{d}s = \widehat{K}(T\omega)e^{-i\omega t_0},
\end{equation}
where $\widehat{K}$ denotes the Fourier transform of $K$.
Thus the signal averaged over finite time still oscillates at frequency $\omega$, but its complex amplitude is multiplied by $\widehat{K}(T\omega)$, which provides a dependency on the averaging duration $T$ as well as an additional dependence on the frequency. By the normalization and integrability of $K$, $\widehat{K}(T\omega)$ approaches $1$ as $T|\omega|\to0$ and $0$ as $T|\omega|\to\infty$. Components with $T|\omega|\ll1$ are therefore retained nearly unchanged, whereas those with $T|\omega|\gg1$ are suppressed, with the transition occurring at frequencies of order $T^{-1}$.

To further illustrate how the shape of averaging windows, as expressed through the kernel $K$, controls frequency retention, we specialize \eqref{eq:spectral-multiplier} to the uniform time average \eqref{eq:centered-uniform-average}. Evaluating \eqref{eq:centered-uniform-average} for $f$ gives
\begin{equation}\label{eq:boxcar-sinc}
(A_T f)(t_0) = \frac{1}{T}\int_{-T/2}^{T/2}e^{-i\omega(t_0+s)}\mathrm{d}s = \operatorname{sinc}\left(\frac{\omega T}{2}\right)e^{-i\omega t_0},
\end{equation}
where $\operatorname{sinc}(x)\defeq \sin(x)/x$, with $\operatorname{sinc}(0)\defeq 1$ (Appendix~\ref{app:boxcar-sinc-response}). Here the uniform time average multiplies the oscillation's complex amplitude by $\operatorname{sinc}(\omega T/2)$, which is not monotone. It vanishes whenever the window contains an integer number of periods but is nonzero between successive zeros, with a sign that alternates from one interval to the next. Thus a uniform average does not necessarily simply suppress higher frequency oscillations more; instead, the contribution of an oscillation with a particular frequency exactly cancels over the averaging window if its cycles fit between the two window endpoints. If a cycle is incomplete between the chosen endpoints, there is a residual contribution associated with that oscillation that can have either sign. Physically, a negative residual shifts the retained oscillation by $\pi$, so a typical (uniform-weighted) time average can effectively alter not only the amplitude but the phase of the oscillation.

If we instead consider averages that do not weight all times equally, the amplitude retained from increasingly rapid oscillations can have a more rapidly decaying envelope than that associated with a uniform average. For a fixed nonzero frequency with $T|\omega|\gg1$, compactly supported $C^\infty$ kernels give $O((T|\omega|)^{-n})$ decay for $n\in\mathbb{Z}^+$ \citep{hautWingate2014asymptoticParallel}, whereas Gaussian kernels give $O(\exp(-(T\omega)^2/2))$ decay \citep{engquistTsai2005HMMStiffODE}. Exponential filters provide a practical compromise by damping high frequencies more smoothly than a uniform average while allowing recursive implementation in time \citep{minzbakerkafiabadvanneste2025}. In contrast to the uniform time average, the standard Gaussian and causal exponential filters have response magnitudes $e^{-(T\omega)^2/2}$ and $\left(1+(T\omega)^2\right)^{-1/2}$, respectively, both of which decrease monotonically as $T|\omega|$ increases. Hence the relative weighting set by the kernel governs how strongly rapid oscillations are suppressed, whereas the effective averaging duration sets the slowest oscillations distinguishable from the mean. However, stronger suppression of frequencies far from zero does not necessarily improve the fidelity of the finite-window averaged dynamics to the original dynamics \citep{bright2011movingAverages}.

We now extend the infinite-time averaging result in \eqref{eq:exact-wave-filter}, which uses uniform weighting, to the kernel-weighted form \eqref{eq:kernel-average}, considering finite Fourier sums. Write
\begin{equation}\label{eq:finite-fourier-sum}
g(t) = \hat g_0 + \sum_{n=1}^{N} \hat g_n e^{-i\omega_n t},
\end{equation}
where $\omega_n\neq0$. Because $A_T$ is linear, applying \eqref{eq:spectral-multiplier} to each component of \eqref{eq:finite-fourier-sum} gives
\begin{equation}\label{eq:finite-fourier-average}
(A_T g)(t_0) = \hat g_0 + \sum_{n=1}^{N}\hat g_n\widehat{K}(T\omega_n)e^{-i\omega_n t_0}.
\end{equation}
Thus each nonzero-frequency component is weighted by $\widehat{K}(T\omega_n)$ at finite $T$.
For each fixed nonzero $\omega_n$, $|T\omega_n|\to\infty$, so the sum in \eqref{eq:finite-fourier-average} vanishes as $T\to\infty$. We denote this infinite-time limit by $A_\infty$:
\begin{equation}\label{eq:infinite-time-limit}
(A_\infty g)(t_0) \defeq \lim_{T\to\infty}(A_T g)(t_0) = \hat g_0.
\end{equation}
For a finite Fourier sum of the form \eqref{eq:finite-fourier-sum}, $\hat g_0$ is simply its time-independent, or zero-frequency, component. If $g(t)$ describes the evolution of an instantaneous state $w_0$ under the fast linear dynamics, $g(t)=e^{-t\mathcal L_{\mathrm F}}w_0$, then \eqref{eq:exact-wave-filter} identifies $\hat g_0$ with $Pw_0$. 

\begin{figure}
  \centering
  \includegraphics[width=\linewidth]{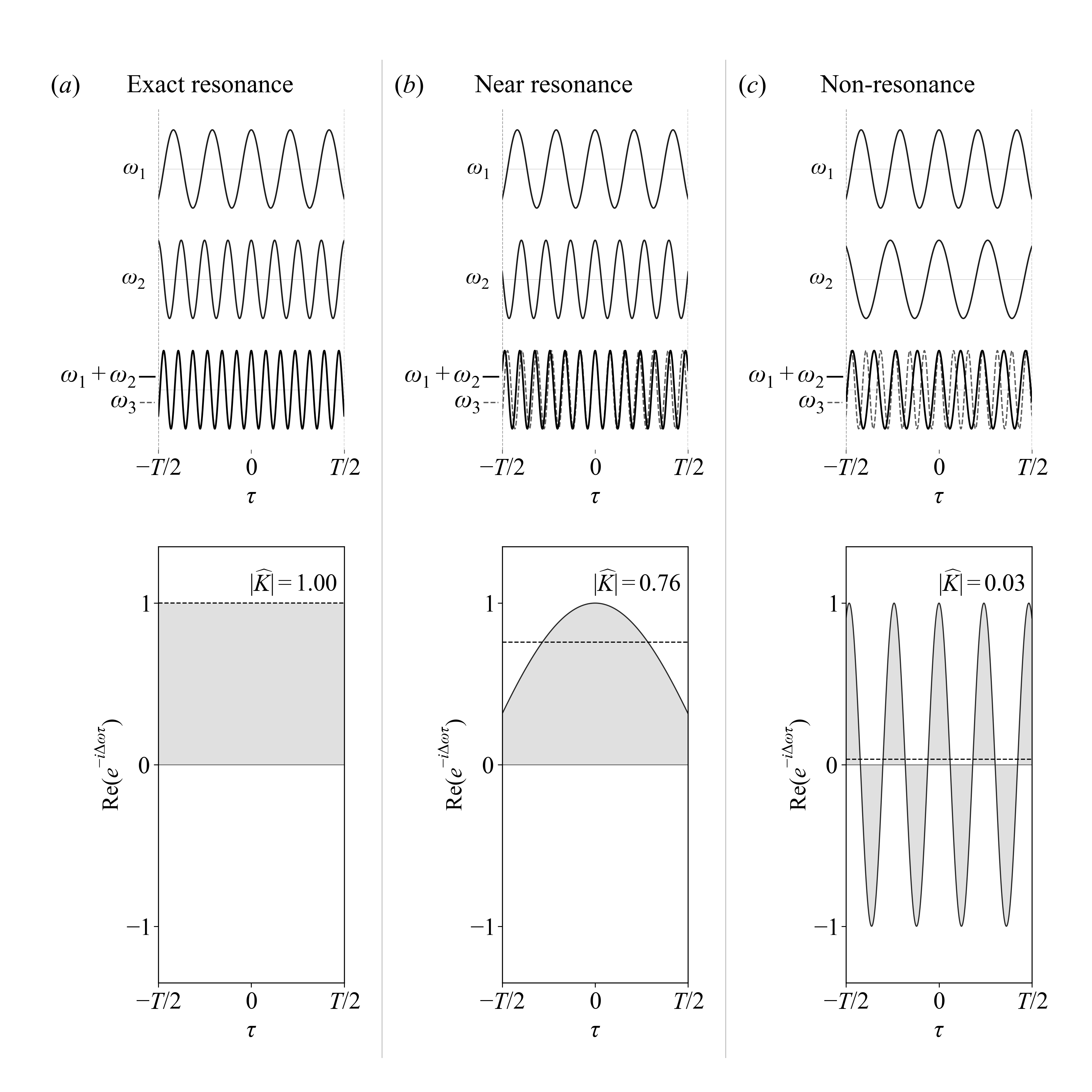}
  \caption{Three regimes of a triad interaction between modes: exact resonance, near resonance, and non-resonance. Top row: two modes at frequencies $\omega_1$ (fixed across panels) and $\omega_2$ (varied) interact quadratically, driving a third mode. This driving oscillates at frequency $\omega_1+\omega_2$ (solid). For comparison, the dashed curve shows how the third mode would oscillate on its own, at its natural linear frequency $\omega_3$. In~($a$), $\omega_2$ is chosen so that $\omega_1 + \omega_2$ matches $\omega_3$ exactly; in~($b$), it is chosen so that the interaction is slightly detuned and the forcing and third mode slowly drift apart; in~($c$), it is chosen so that the interaction is far from resonance and the forcing and third mode have visibly different frequencies. Bottom row: the relative phase factor $\mathrm{Re}(e^{-i\Delta\omega\tau})$ over a centered uniform window of length $T$ in fast time $\tau$. Here $\Delta\omega = \omega_1 + \omega_2 - \omega_3$ is the signed frequency mismatch, and the dashed line marks its window mean $\widehat{K}(T\Delta\omega)$, whose magnitude measures how much of the interaction is retained and decreases from $|\widehat{K}| = 1$ (fully retained) to $0.76$ (mostly retained) to $0.03$ (largely cancelled).}
  \label{fig:triad-detuning}
\end{figure}

We can now use the frequency response to quantify how finite averaging changes the resonance selection in \eqref{eq:averaged-equation}. A finite window retains some detuned interactions that the infinite-time average removes. To see how, consider two modes with wavenumbers $\bm{k}_1$ and $\bm{k}_2$ and associated linear fast frequencies $\omega_1$ and $\omega_2$, respectively. Their quadratic interaction drives the mode at $\bm{k}_3=\bm{k}_1+\bm{k}_2$ at the combined frequency $\omega_1+\omega_2$. This third mode has a natural linear frequency $\omega_3$, defined even when its amplitude is zero. The forcing accumulates phase relative to that mode at the detuning $\Delta\omega\defeq\omega_1+\omega_2-\omega_3$. For the quadratic interactions in \eqref{eq:averaged-equation}, the signal to be averaged in fast time is their relative phase, $f(\tau)\defeq e^{-i\Delta\omega\tau}$, which is identically unity at exact resonance, where $\Delta\omega=0$. We average this signal over a fast-time window of length $T$ centered at $\tau_0\defeq t_0/\varepsilon$. Applying \eqref{eq:spectral-multiplier} at frequency $\Delta\omega$ gives
\begin{equation}\label{eq:near-matching-general}
(A_T f)(\tau_0) = \int_{\mathbb{R}}K_T(s)e^{-i\Delta\omega(\tau_0+s)}\mathrm{d}s = \widehat{K}\big(T\Delta\omega\big)e^{-i\Delta\omega\tau_0}.
\end{equation}
Specializing \eqref{eq:near-matching-general} to the uniform time average and using the sinc response \eqref{eq:boxcar-sinc} gives
\begin{equation}\label{eq:near-matching-sinc}
(A_T f)(\tau_0) = \frac{1}{T}\int_{-T/2}^{T/2}e^{-i\Delta\omega(\tau_0+s)}\mathrm{d}s = \operatorname{sinc}\Big(\tfrac{\Delta\omega T}{2}\Big)e^{-i\Delta\omega\tau_0}.
\end{equation}
While we saw earlier that a finite average can retain slowly varying contributions to a field, here the slowly varying quantity is the relative phase of an interaction between fast waves, and hence near-resonant interactions are retained. Indeed, the weighting factor $\operatorname{sinc}(T\Delta\omega/2)$ in \eqref{eq:near-matching-sinc} remains appreciable when $|T\Delta\omega|\lesssim1$. The corresponding near-resonant band has width $O(T^{-1})$, so a fast-time window on the scale $T\sim\varepsilon^{-1}$ retains detunings $\Delta\omega=O(\varepsilon)$. As $T\to\infty$, this band narrows to exact resonance, $\Delta\omega=0$, reproducing the zero-frequency selection in \eqref{eq:infinite-time-limit}.  The nonzero-detuning interactions retained in this band are quasi-resonant triads in the same sense as those treated in analyses of rotating, stratified flows \citep[Theorem~3.1]{babinmahalovnicolaenko2002}.

On this basis, we can interpret the cancellation of oscillations \citep[e.g.][]{schochet1994} through the phase accumulated by the forcing produced by the two interacting modes relative to the mode it drives over the averaging window. Figure~\ref{fig:triad-detuning} illustrates this cancellation for the triad described above: in each case the lower plot is the relative phase factor $\mathrm{Re}(e^{-i\Delta\omega\tau})$ across the window, and the fraction the average retains is its window mean $\widehat{K}(T\Delta\omega)$. At exact resonance the relative phase does not oscillate over the window, so nothing cancels and the entire contribution to the dynamics is retained. At small detuning, where the interaction is near-resonant, the relative phase completes less than a full oscillation over the window, so it cannot average to zero, and most but not all of the contribution remains. At large detuning it runs through many oscillations whose positive and negative parts nearly cancel, leaving almost none of the contribution behind.

\subsection{Closure and accuracy under finite-window averaging}\label{sec:finite-window-closure-accuracy}

To complete the comparison of finite-time and infinite-time averages, we turn from the triad interactions they retain to the closure and accuracy of the corresponding averaged equations. Classical averaging gives a closed amplitude equation containing only exact resonances. To obtain a finite-window equation, we replace the infinite fast-time average in \eqref{eq:averaged-equation} with $A_T$, evaluated at the fast phase $\tau+s$ as $s$ ranges over the averaging window. We hold the amplitude fixed at its value at time $t$ and integrate over $s$ to determine its averaged rate of change at that time. Finite-window averaging of the transformed nonlinear terms is also used in numerical methods for oscillatory PDEs \citep{hautWingate2014asymptoticParallel,peddleHautWingate2019parareal}. The resulting amplitude $\overline{\bm{w}}_{K,T}$ (the subscripts indicate its dependence on the averaging kernel $K_T$) is governed by
\begin{equation}\label{eq:finite-window-closed-evolution}
\begin{split}
\frac{\partial \overline{\bm{w}}_{K,T}}{\partial t} + \int_{\mathbb{R}} K_T(s)e^{(\tau+s)\mathcal{L}_{\mathrm F}}\Bigl(&\mathcal{L}_S(e^{-(\tau+s)\mathcal{L}_{\mathrm F}}\overline{\bm{w}}_{K,T}) + \mathcal{B}(e^{-(\tau+s)\mathcal{L}_{\mathrm F}}\overline{\bm{w}}_{K,T},e^{-(\tau+s)\mathcal{L}_{\mathrm F}}\overline{\bm{w}}_{K,T})\\
&-\mathcal{D}(e^{-(\tau+s)\mathcal{L}_{\mathrm F}}\overline{\bm{w}}_{K,T})\Bigr)\mathrm{d}s = 0.
\end{split}
\end{equation}
To separate the contributions from individual wave interactions, we can again expand the quadratic term in \eqref{eq:finite-window-closed-evolution} in modes, as detailed in Appendix~\ref{app:closed-finite-window-dynamics}, and apply \eqref{eq:spectral-multiplier} to each interaction's relative phase $e^{-i\Delta\omega(\tau+s)}$, so integration against $K_T$ yields $e^{-i\Delta\omega\tau}\widehat{K}(T\Delta\omega)$. The corresponding infinite-time average instead assigns the weight $\mathbf{1}_{\Delta\omega=0}$, so only interactions with zero detuning enter the exact averaged dynamics (Appendix~\ref{app:closed-finite-window-dynamics}). The significance of this formulation is that it allows us to quantify how changing the averaging window alters the contributions of detuned interactions to the averaged dynamics. Although the finite window retains interactions with $\Delta\omega\neq0$, \eqref{eq:finite-window-closed-evolution} remains closed because the integral is fully determined by the amplitude $\overline{\bm{w}}_{K,T}(t)$, the prescribed fast evolution, and the chosen kernel. By contrast, an equation for $A_T\bm{w}$ (the finite-window average of the evolving state itself, rather than the solution of the finite-window averaged equation) would contain averages of nonlinear products that are not determined by $A_T\bm{w}$ alone.

While our finite-window generalization of classical averaging shows how much of each additional wave interaction is retained in the (finite-time) averaged dynamics, averaging is often used to obtain a reduced description of a selected dynamical feature, so one may also need to know how faithfully that description represents the dynamics of interest. Using classical averaging theory, one can quantify this fidelity through an error bound comparing identically initialized solutions of the original and first-order infinite-time averaged equations; for periodic problems under standard regularity and boundedness assumptions, their difference is $O(\varepsilon)$ over intervals of length $O(1)$ in the slow time $t$ \citep[theorem~2.8.1]{sandersVerhulstMurdock2007averaging}. To apply this result to an oscillatory problem, one first maps out the known fast motion, averages the equation governing the resulting slow variables, and subsequently maps the solution of the averaged equation back into the original variables \citep[sections~1.6--1.7]{sandersVerhulstMurdock2007averaging}. As a concrete example, directly averaging the time-dependent damping coefficient in a periodically damped oscillator gives an $O(1)$ error in the predicted amplitude decay over the slow timescale, whereas averaging the equations for its slowly varying amplitude and phase gives an $O(\varepsilon)$ approximation \citep[section~2.7]{sandersVerhulstMurdock2007averaging}. If the near-inertial amplitude is the quantity of interest, averaging the velocity without first mapping out the inertial oscillation recovers the background $U$ with $O(\varepsilon)$ error but discards the $O(1)$ amplitude $A$, whereas performing the mapping before averaging recovers the slowly varying amplitude $A$ with $O(\varepsilon)$ error, as shown in \eqref{eq:intro-niw-direct-average}--\eqref{eq:intro-niw-mapped-average}. Thus the fidelity of the resulting reduced description depends not only on the window length and the relative weighting within it but also on the variables in which averaging is performed.

\section{Nearly-slow drift under finite-time averaging}\label{sec:drift}

The previous section showed that finite-window averaging retains near-resonant interactions that exact averaging removes. A natural quantity in which to examine the dynamical effect of this retained content is drift: it is tied to the reference frame in which the average is taken and, arising from nonlinear interactions in the oscillatory motion, is sensitive to which interactions the average retains. Under Lagrangian averaging, a wave field can produce Stokes drift, an example of wave-induced drift arising from the quadratic coupling between the induced displacement and the gradient of the oscillatory velocity \citep{andrewsmcintyre1978a,buhlermcintyre1998}. To examine this drift, we follow the Eulerian route of \citet{wagneryoung2015}, who derive a wave-averaged QG equation from the Boussinesq equations and recover the Lagrangian-mean transport structure, including its quadratic drift correction, of \citet{buhlermcintyre1998}.

The Stokes drift calculation in \S\S\ref{sec:hydrostatic-fast-dynamics}--\ref{sec:nearly-slow-drift} proceeds in two main steps: first, we use the exact projection $P$ to separate balanced motion from the wave motion used to construct the velocity and displacement. Second, we average their quadratic correlation. This average is first taken over infinite time. However, we do not apply the exponential mapping in~\eqref{eq:exact-wave-filter} as an additional transformation before averaging. While an infinite-time average of an arbitrary history need not realize the exact projection $P$ associated with the selected fast dynamics, here that evolution is already contained in the velocity and displacement constructed from the wave component in the first step, appearing explicitly as oscillatory phase factors in their modal expansions. For periodic wave motion, this infinite-time average is equivalent to the average over a complete wave period used in the classical Stokes drift calculation. For a general wave field, the infinite average extends averaging over complete wave periods by removing fast-time oscillations from the correlation between displacement and velocity gradient at fixed slow time. We subsequently compare this drift with that obtained via finite averaging instead, to determine whether the retained near-resonant interactions contribute additional transport. In this way, the calculation connects the drift associated with Lagrangian averaging to the resonance selection of classical averaging.

We can see the connection between Stokes drift and wave interactions in the classical calculation of \citet{stokes1847oscillatoryWaves} for a small-amplitude deep-water wave. In this case, the surface elevation $\eta=a\cos(kx-\omega t)$ oscillates rapidly in physical time $t$, with $\omega$ satisfying the deep-water dispersion relation $\omega^2=gk$. To leading order in the small wave steepness $ka\ll1$, one may solve the linearized irrotational free-surface problem to obtain the velocity field
\begin{equation}\label{eq:classical-stokes-ex-1}
(u,w)^\top=a\omega e^{kz}(\cos(kx-\omega t),\sin(kx-\omega t))^\top,
\end{equation}
with $z$ measured upwards from the undisturbed surface. Integrating this velocity at fixed $(x,z)$ gives the zero-mean oscillatory displacement
\begin{equation}\label{eq:classical-stokes-ex-2}
(\xi,\zeta)^\top=ae^{kz}(-\sin(kx-\omega t),\cos(kx-\omega t))^\top,
\end{equation}
where $(\partial_t\xi,\partial_t\zeta)^\top=(u,w)^\top$.

We now calculate the Stokes drift, which can be defined as the difference between the Lagrangian- and Eulerian-mean velocities. Here `mean' refers to a uniform time average over one wave period $T=2\pi/\omega$, denoted $\langle\cdot\rangle_T$. Averaging the velocity at a fixed position gives the Eulerian mean,
\begin{equation}\label{eq:classical-stokes-ex-3}
\left\langle u(x,z,t)\right\rangle_T=a\omega e^{kz}\left\langle\cos(kx-\omega t)\right\rangle_T=0.
\end{equation}
To calculate the Lagrangian mean, we need the velocity at the displaced particle position $(x+\xi,z+\zeta)$. Following Stokes' calculation, we Taylor expand this velocity about $(x,z)$ in the small displacement to obtain
\begin{equation}\label{eq:classical-stokes-ex-4}
u(x+\xi,z+\zeta,t)=u(x,z,t)+\xi\partial_xu+\zeta\partial_zu+O(k^2a^3\omega).
\end{equation}
Averaging the velocity at the displaced particle position gives the Lagrangian mean,
\begin{equation}\label{eq:classical-stokes-ex-5}
\left\langle u(x+\xi,z+\zeta,t)\right\rangle_T=\left\langle\xi\partial_xu+\zeta\partial_zu\right\rangle_T+O(k^2a^3\omega).
\end{equation}
The difference between \eqref{eq:classical-stokes-ex-5} and \eqref{eq:classical-stokes-ex-3} gives the (leading-order) Stokes drift,
\begin{align}
u^S &=\left\langle\xi\partial_xu+\zeta\partial_zu\right\rangle_T \notag\\
&=ka^2\omega e^{2kz}\left\langle\sin^2(kx-\omega t)+\cos^2(kx-\omega t)\right\rangle_T \notag\\
&=ka^2\omega e^{2kz}. \label{eq:classical-stokes-ex-6}
\end{align}
Notice that the Stokes drift \eqref{eq:classical-stokes-ex-6} is a slow quantity (here, steady even in physical time): it is independent of $t$ and does not oscillate, as one would intuitively expect of a non-wave contribution. However, it is formed from fast--fast interactions; both the displacement $\xi=-ae^{kz}\sin(kx-\omega t)$ and the velocity gradient $\partial_xu=-ka\omega e^{kz}\sin(kx-\omega t)$ are functions of $t$ and oscillate rapidly, as do the constituents of $\zeta\partial_zu$, which together comprise the steady drift. Thus Stokes drift may be viewed as the slow velocity produced by two interacting fast wave fields, associated with the displacement and velocity, respectively (or, equivalently, by a single fast wave interacting with itself). In this section, we will see that Stokes drift arises from fast--fast--slow interactions in general flows, not only in the case of a single deep water surface gravity wave.

\subsection{The fast dynamics and their modal structure}\label{sec:hydrostatic-fast-dynamics}

The Stokes drift considered here is built from the fast wave velocity and the particle displacement obtained by integrating that velocity in fast time, so we begin by specifying the fast hydrostatic motion and isolating its oscillatory component from the balanced mode. Under the QG ordering of \S\ref{sec:governing-fast-structure}, this physical fast motion evolves as $e^{-\tau\mathcal L_{\mathrm F}}$, where $\mathcal L_{\mathrm F}$ is the normalized $O(1)$ operator in \eqref{eq:abstract-system}, with the common $\varepsilon^{-1}$ factor extracted.

Two orderings enter the construction. First, the scaling of the governing equations selects which motion is treated as fast, and so fixes the oscillations that define the exact and finite wave averages. Second, as in the preceding derivation of Stokes drift, we expand in the small wave displacement. Writing
\begin{equation}\label{eq:rossby-numbers}
\Ro_w\defeq \frac{U_w}{fL}, \qquad \Ro_b\defeq \frac{U_b}{fL}
\end{equation}
for the wave and balanced Rossby numbers, respectively, the ordering used by \citet{wagneryoung2015} takes $\Ro_w$ as the small parameter and sets $\Ro_b\sim\Ro_w^2$. For fixed modes with $O(1)$ nondimensional wavenumber and frequency, the wave displacement scaled by $L$ is $O(\Ro_w)$, so the quadratic wave drift is $O(\Ro_w^2)$, comparable to the balanced transport $O(\Ro_b)$, with both velocities expressed in units of $fL$. If instead $\Ro_b\sim\Ro_w$, as in the single-amplitude setup of \S\ref{sec:governing-fast-structure}, the same drift is one order smaller than the leading balanced transport and would enter as one of several contributions to the averaged mean-flow equation at the next order.

As in \citet{wagneryoung2015}, we specialize to the hydrostatic system by replacing the vertical component of the momentum equation in \eqref{eq:momentum-em98} with hydrostatic balance $\partial_z p=-\Gamma\rho$, using $p=\Eu\phi$ and the parameters defined in \eqref{eq:boussinesq-dimensionless-parameters}. On Fourier modes with vertical wavenumber $m\neq0$, where $\partial_z^{-1}$ is the multiplier $(im)^{-1}$, we use hydrostatic balance and incompressibility to write the fast dynamics for $\bm{w}_H\defeq(u,v,\rho)^\top$ (with $H$ denoting `hydrostatic' or `horizontal', since the vertical velocity $w$ has been eliminated) as
\begin{subequations}\label{eq:LF-hydrostatic}
\begin{align}
\partial_\tau\bm{w}_H+\mathcal{L}_{\mathrm F}\bm{w}_H&=0, \label{eq:LF-hydrostatic-evolution}\\
\mathcal{L}_{\mathrm F}\begin{pmatrix}u\\ v\\ \rho\end{pmatrix} &= \begin{pmatrix} -F v-\partial_x\partial_z^{-1}\rho\\ F u-\partial_y\partial_z^{-1}\rho\\ \partial_z^{-1}(\partial_x u+\partial_y v) \end{pmatrix}. \label{eq:LF-hydrostatic-action}
\end{align}
\end{subequations}
On a periodic domain and for Fourier modes with wavevector $\bm{k}=(k,l,m)$, the Fourier symbol of the fast operator $\mathcal L_{\mathrm F}$ is
\begin{equation}\label{eq:LF-fourier}
\mathcal{L}_{\mathrm F}(\bm{k})= \begin{pmatrix} 0 & -F & -\dfrac{k}{m}\\[2ex] F & 0 & -\dfrac{l}{m}\\[2ex] \dfrac{k}{m} & \dfrac{l}{m} & 0 \end{pmatrix},
\end{equation}
where $m\neq0$. The eigenvalues of~\eqref{eq:LF-fourier} are $i\omega^{(\beta)}(\bm{k})$, $\beta\in\{-1,0,1\}$, where
\begin{equation}\label{eq:dispersion-hydrostatic}
\omega^{(\beta)}(\bm{k})\defeq\beta\omega(\bm{k}), \qquad \omega(\bm{k})\defeq\left(F^2+\frac{k^2+l^2}{m^2}\right)^{1/2}.
\end{equation}
The zero frequency~$\omega^{(0)}$ corresponds to the balanced (geostrophic) mode, and the two nonzero frequencies~$\omega^{(\pm1)}$ to the hydrostatic inertia--gravity modes.

In the classical example \eqref{eq:classical-stokes-ex-6}, the oscillatory velocity had already been isolated \eqref{eq:classical-stokes-ex-1}. Here, isolating the oscillatory motion requires us first to decompose the hydrostatic state into balanced and wave components. To separate these components, we use the exact projection $P$ in \eqref{eq:exact-wave-filter}, which selects the zero-frequency balanced mode at each wavevector $\bm k$.

The leading-order balance~\eqref{eq:LF-hydrostatic-evolution} has the solution form~\eqref{eq:leading-order-solution}, with a general slowly varying amplitude $\overline{\bm{w}}_H$. Because Stokes drift is generated by the oscillatory motion, we decompose this amplitude into its balanced and wave components:
\begin{equation}\label{eq:amplitude-decomposition}
\overline{\bm{w}}_H = \overline{\bm{w}}_{\mathrm{bal}} + \overline{\bm{w}}_{\mathrm{wave}}, \qquad \overline{\bm{w}}_{\mathrm{bal}}\defeq P\overline{\bm{w}}_H, \qquad \overline{\bm{w}}_{\mathrm{wave}}\defeq(I-P)\overline{\bm{w}}_H.
\end{equation}
Here the wave subspace is $\operatorname{ran}(I-P)=\ker P$; equivalently, at each wavevector $\bm k$, its Fourier coefficients lie in the span of the $\beta=\pm1$ eigenvectors of $\mathcal L_{\mathrm F}(\bm k)$. Hence $P\overline{\bm w}_{\mathrm{wave}}=0$.

\subsection{Constructing the Stokes drift}

To construct the Stokes drift, we first recover the full velocity from the reduced hydrostatic state through incompressibility by defining a reconstruction map~$\mathcal{I}_{\bm{u}}$:
\begin{equation}\label{eq:velocity-reconstruction}
\mathcal{I}_{\bm{u}}\begin{pmatrix}u\\ v\\ \rho\end{pmatrix} \defeq \begin{pmatrix} u\\ v\\ -\partial_z^{-1}(\partial_x u+\partial_y v) \end{pmatrix}, \qquad \widehat{\mathcal{I}_{\bm{u}}\bm{w}_H}(\bm{k}) = \begin{pmatrix} \widehat u\\ \widehat v\\ -\dfrac{k\widehat u+l\widehat v}{m} \end{pmatrix},
\end{equation}
where $m\neq0$. The fast wave field and the corresponding physical velocity are then
\begin{equation}\label{eq:fast-field}
\bm{w}_{\mathrm{fast}}(\bm{x},\tau,t) \defeq e^{-\tau\mathcal{L}_{\mathrm F}}\overline{\bm{w}}_{\mathrm{wave}}(\bm{x},t), \qquad \bm{u}_{\mathrm{fast}}(\bm{x},\tau,t) \defeq \mathcal{I}_{\bm{u}}\bm{w}_{\mathrm{fast}}(\bm{x},\tau,t).
\end{equation}
Because $P\overline{\bm{w}}_{\mathrm{wave}}=0$, the fast orbit contains only the nonzero-frequency wave modes, on which $\mathcal{L}_{\mathrm F}^{-1}$ is defined. Consequently, $A_\infty\bm{u}_{\mathrm{fast}}=\mathcal{I}_{\bm{u}}P\overline{\bm{w}}_{\mathrm{wave}}=0$, with $A_\infty$ as in~\eqref{eq:infinite-time-limit}, and the reconstructed fast velocity admits a bounded fast-time antiderivative. Since $\tau=t/\varepsilon$, converting the fast-time derivative to the physical-time derivative introduces a factor $\varepsilon^{-1}$. The oscillatory particle displacement therefore satisfies
\begin{equation}\label{eq:displacement-condition}
\partial_\tau\bm{\xi}(\bm{x},\tau,t)=\varepsilon\bm{u}_{\mathrm{fast}}(\bm{x},\tau,t), \qquad (A_\infty\bm{\xi})(\bm{x},t)=0.
\end{equation}

Substituting the fast velocity from~\eqref{eq:fast-field} into the displacement equation~\eqref{eq:displacement-condition}, we integrate in $\tau$ at fixed slow time. The wave amplitude and reconstruction map are independent of $\tau$, so only the exponential is integrated, introducing the factor $-\mathcal L_{\mathrm F}^{-1}$ and an additive $\tau$-independent integration constant. The oscillatory term already has zero exact fast mean, so the mean condition in~\eqref{eq:displacement-condition} sets this constant to zero, giving
\begin{equation}\label{eq:displacement}
\bm{\xi}(\bm{x},\tau,t) = -\varepsilon\mathcal{I}_{\bm{u}}e^{-\tau\mathcal{L}_{\mathrm F}}\mathcal{L}_{\mathrm F}^{-1} \overline{\bm{w}}_{\mathrm{wave}}(\bm{x},t).
\end{equation}
Hence, integrating the fast-wave velocity allows the associated particle displacement to be expressed in terms of the slowly varying wave amplitude. Here $\overline{\bm w}_{\mathrm{wave}}(\bm x,t)$ carries the spatial structure and slow-time variation of the waves, while $e^{-\tau\mathcal L_{\mathrm F}}$ generates their fast oscillations. Applying $\mathcal I_{\bm u}$ gives the full particle displacement, preserving its horizontal components and recovering its vertical component through incompressibility. The resulting displacement \eqref{eq:displacement} enters the correlation with the wave-velocity gradient whose fast-time average gives the Stokes drift, as in the classical example \eqref{eq:classical-stokes-ex-6}.

Generalizing the expansion in \eqref{eq:classical-stokes-ex-4}, we evaluate the advecting velocity at the displaced particle position $\bm{x}+\bm{\xi}$ and expand to first order in $\bm{\xi}$:
\begin{equation}\label{eq:lifted-velocity-expansion}
\bm{u}(\bm{x}+\bm{\xi},\tau,t) = \bm{u}_{\mathrm{slow}} +\bm{u}_{\mathrm{fast}} +(\bm{\xi}\cdot\nabla)\bm{u}_{\mathrm{fast}} +(\bm{\xi}\cdot\nabla)\bm{u}_{\mathrm{slow}} +\cdots .
\end{equation}
This first-order truncation uses the small-displacement ordering $|\bm{\xi}\cdot\nabla|=O(\Ro_w)\ll1$, with $\Ro_w$ defined in~\eqref{eq:rossby-numbers}, so the omitted terms are higher order in $\Ro_w$. Since $\bm{u}_{\mathrm{slow}}$ has zero fast frequency, the slow--fast term $(\bm{\xi}\cdot\nabla)\bm{u}_{\mathrm{slow}}$ retains the nonzero wave frequencies carried by $\bm{\xi}$ and is therefore removed by $A_\infty$. A finite window may retain part of this term, but the averaged term remains a fast oscillation and contributes neither to the zero-frequency exact drift nor to the nearly-slow drift. By contrast, the fast--fast term $(\bm{\xi}\cdot\nabla)\bm{u}_{\mathrm{fast}}$ contains wave--wave correlations in which the two fast frequencies can cancel exactly or nearly. We therefore isolate this term to compare the corresponding drifts under exact and finite-time averaging:
\begin{equation}\label{eq:drift-diagnostic}
\bm{u}_A^S(\bm{x},\tau,t) \defeq \bigl(A((\bm{\xi}\cdot\nabla)\bm{u}_{\mathrm{fast}})\bigr)(\bm{x},\tau,t), \qquad A\in\{A_\infty, A_T\}.
\end{equation}
To isolate the effect of the averaging window, we use the same displacement $\bm\xi$, chosen so that $A_\infty\bm\xi=0$, for both averages at the same reference position $\bm{x}$ and vary only the operator applied to the quadratic correction. Computing the classical Stokes drift at this order amounts to averaging $(\bm\xi\cdot\nabla)\bm u_{\mathrm{fast}}$, the velocity correction induced by the oscillatory particle displacement. For a wave field that is periodic at fixed slow time, applying $A_\infty$ is equivalent to the familiar averaging over complete wave cycles, giving the classical Stokes drift $\bm{u}_{A_\infty}^S$. As we show below, finite averaging gives a distinct, window-dependent drift $\bm{u}_{A_T}^S$, which can differ from $\bm{u}_{A_\infty}^S$ by the contribution we term the \emph{nearly-slow drift}.

To compare the drift under exact and finite averaging, we express the displacement and velocity as sums of wave modes and track the combined frequency of each resulting pair. We label each nonzero-frequency wave mode by its wavevector and frequency sign, $a=(\bm k_a,\beta_a)$, collect these labels in $\Lambda$, and write the signed frequency of mode $a$ as $\omega_a\defeq\omega^{(\beta_a)}(\bm k_a)$. Expanding $\overline{\bm w}_{\mathrm{wave}}$ in these modes and evolving each with its fast phase gives the reduced wave field. Applying the velocity reconstruction map~\eqref{eq:velocity-reconstruction} gives $\bm u_{\mathrm{fast}}$, and integrating the reconstructed velocity in fast time according to~\eqref{eq:displacement-condition} gives $\bm\xi$:
\begin{equation}\label{eq:mode-amplitudes}
\bm{u}_{\mathrm{fast}} = \sum_{a\in\Lambda} \bm{U}_a e^{-i\omega_a\tau}, \qquad \bm{\xi} = -\varepsilon\sum_{a\in\Lambda} \frac{\bm{U}_a}{i\omega_a}e^{-i\omega_a\tau}.
\end{equation}
Here $\bm{U}_a(\bm{x},t) = \sigma_{\bm{k}_a}^{(\beta_a)}(t)e^{i\bm{k}_a\cdot\bm{x}}\mathcal{I}_{\bm{u}}r_{\bm{k}_a}^{(\beta_a)}$ is the spatially dependent complex velocity amplitude of mode $a$, where $r_{\bm{k}}^{(\beta)}$ is the corresponding eigenvector and $\sigma_{\bm{k}}^{(\beta)}(t)$ are the modal coefficients in the eigenbasis expansion of $\overline{\bm{w}}_{\mathrm{wave}}$.

The quantity averaged in \eqref{eq:drift-diagnostic}, $(\bm\xi\cdot\nabla)\bm u_{\mathrm{fast}}$, pairs each displacement mode $a$ with the gradient of each velocity mode $b$. Because the displacement and velocity phases multiply in the drift integrand, the resulting wave--wave correlation oscillates at the signed sum of their frequencies. Substituting the modal expansions~\eqref{eq:mode-amplitudes} into $(\bm{\xi}\cdot\nabla)\bm{u}_{\mathrm{fast}}$ gives
\begin{equation}\label{eq:product-pairing}
(\bm{\xi}\cdot\nabla)\bm{u}_{\mathrm{fast}} = \sum_{a,b\in\Lambda} \bm{C}_{ab}e^{-i\Delta\omega_{ab}\tau},
\end{equation}
where the pairwise drift amplitude and frequency mismatch are defined, respectively, by
\begin{equation}\label{eq:pairwise-drift-amplitude}
\bm{C}_{ab}(\bm{x},t)\defeq -\varepsilon\left(\frac{\bm{U}_a}{i\omega_a}\cdot\nabla\right)\bm{U}_b, \qquad \Delta\omega_{ab}\defeq\omega_a+\omega_b.
\end{equation}
The modal sums include conjugate mode pairs, so the resulting physical drift is real.
In the notation of \S\ref{sec:exact-and-finite}, $\Delta\omega_{ab}$ measures how far the interaction between displacement mode $a$ and velocity mode $b$ is from producing a zero-fast-frequency drift contribution.
Because the fast-time average holds $\bm x$ and $t$ fixed, the pairwise amplitude $\bm C_{ab}(\bm x,t)$ can be taken outside the average, leaving only the phase factor $e^{-i\Delta\omega_{ab}\tau}$ to be averaged. In a time record sampled at fixed $\bm x$, the amplitudes evolve along with the fast phases, so the average calculated here represents the record's average accurately when $\bm C_{ab}$ changes negligibly over the window of duration $\varepsilon T$ in $t$. For $O(1)$ fast frequencies and amplitudes evolving on the advective timescale, $1\ll T\ll\varepsilon^{-1}$ permits many wave cycles with little amplitude change, whereas windows with $T\sim\varepsilon^{-1}$ generally require accounting for amplitude evolution within the record.

\subsection{Nearly-slow drift}\label{sec:nearly-slow-drift}

The modal expansion now allows us to compare how exact and finite averaging retain different transport contributions from the same wave field. By~\eqref{eq:infinite-time-limit}, $A_\infty$ keeps only pairs with $\Delta\omega_{ab}=0$:
\begin{equation}\label{eq:drift-exact}
\bm{u}_{A_\infty}^S = \sum_{\substack{a,b\in\Lambda\\ \Delta\omega_{ab}=0}} \bm{C}_{ab}.
\end{equation}
Hence the retained fast-mode pairs generate a zero-fast-frequency drift and satisfy $\Delta\omega_{ab}=\omega_a+\omega_b=0$, the frequency condition for a fast--fast--slow resonance. Here `slow' refers to the zero fast-time frequency of the transport correction. Under finite-window averaging, each pairwise phase is instead weighted according to its frequency mismatch, as in~\eqref{eq:near-matching-general}:
\begin{equation}\label{eq:drift-finite-window}
\bm{u}_{A_T}^S = \sum_{a,b\in\Lambda} \widehat{K}(T\Delta\omega_{ab}) e^{-i\Delta\omega_{ab}\tau}\bm{C}_{ab}.
\end{equation}
Finite averaging reweights each pair without eliminating its phase factor, so each retained contribution with $\Delta\omega_{ab}\ne0$ continues to oscillate at the combined frequency of the two waves. For a uniform time average,~\eqref{eq:near-matching-sinc} gives $\widehat{K}(T\Delta\omega_{ab})=\operatorname{sinc}(\Delta\omega_{ab}T/2)$, so pairings with~$|\Delta\omega_{ab}|\lesssim 1/T$ are retained with appreciable weight. Thus, the finite-window drift contains near-resonant contributions that are absent from the exact drift defined in~\eqref{eq:drift-exact}.

Because $\widehat K(0)=1$, subtracting the exact drift~\eqref{eq:drift-exact} from the finite-window drift~\eqref{eq:drift-finite-window} cancels the exactly resonant contributions, for which $\Delta\omega_{ab}=0$, leaving
\begin{equation}\label{eq:drift-residual}
\bm{u}^S_{A_T} - \bm{u}^S_{A_\infty} = \sum_{\substack{a,b\in\Lambda\\ \Delta\omega_{ab}\neq0}} \widehat{K}(T\Delta\omega_{ab}) e^{-i\Delta\omega_{ab}\tau}\bm{C}_{ab}.
\end{equation}
Each term in~\eqref{eq:drift-residual} oscillates at the nonzero frequency mismatch~$\Delta\omega_{ab}$, which produces slow variation over the averaging window for near-resonant pairs satisfying $0<|\Delta\omega_{ab}|T\lesssim1$, as illustrated in figure~\ref{fig:triad-detuning}($b$). These near-resonant contributions are therefore nearly slow relative to that window, forming what we call the nearly-slow drift. For $|\Delta\omega_{ab}|T\gg1$, the multiplier can be small but nonzero, as illustrated in figure~\ref{fig:triad-detuning}($c$), so the full difference \eqref{eq:drift-residual} can include attenuated terms that still oscillate rapidly through $e^{-i\Delta\omega_{ab}\tau}$ and hence are not considered part of the nearly-slow drift.

Like the familiar (leading-order) Stokes drift in GLM \citep{andrewsmcintyre1978a} and later wave-averaged treatments \citep{wagneryoung2015,vanneste2022stokes}, the nearly-slow drift has a quadratic structure coupling wave displacement to the gradient of wave velocity. Earlier surface-wave studies identified a zero-frequency contribution to the mean Stokes drift \citep{herterichhasselmann1982} and described contributions produced by waves with slightly different frequencies as infragravity Stokes-drift fluctuations associated with wave groups \citep{herbersjanssen2016}; \citet{smith2006} observed strong variations in Stokes drift during the passage of compact wave groups. Previous studies have connected Stokes drift with resonant interactions in particular settings \citep[e.g.][]{gupta2021}. In deep water, resonant four-wave interactions can generate new propagating waves \citep{phillips1960,longuethiggins1962} and modify the phase speeds of existing waves, for two co-propagating waves by an amount equal to the longer wave's surface Stokes drift \citep{longuethigginsphillips1962}. Here our formulation makes a different, general connection explicit, relating the resonance structure of the drift to the frequency selection imposed by averaging and quantifying how the averaging window weights the contribution from each pair of waves.

Importantly, with the averaging formulation of \S\ref{sec:exact-and-finite}, Stokes drift can now be expressed through resonant wave interactions \eqref{eq:drift-exact}, with near-resonant pairs retained under finite-window averaging giving rise to a distinct nearly-slow drift. Writing the drift in terms of wave interactions clarifies how fast motion produces a slow transport correction: two fast wave modes combine to produce a slow drift contribution when their combined frequency vanishes and, under finite-window averaging, a nearly-slow drift contribution when their combined frequency is small but nonzero. For periodic wave motion, classical Stokes drift is defined by an average over a full wave period, whereas a drift diagnosed from a finite observational or simulation record can include the detuned pairings retained by the window and hence the nearly-slow drift.

\section{Numerical simulations}\label{sec:numerical-simulations}

In this section, we illustrate the preceding analysis and its physical implications through a series of numerical experiments. We first consider a minimal modification of the stationary Rossby wave example illustrated in figure~\ref{fig:andrews-single-wave} that produces a nonzero nearly-slow drift \eqref{eq:drift-residual}: adding a weaker wave at a nearby frequency creates a detuned pairing whose weight in the drift is set by the choice of averaging window. Using a fully nonlinear rotating shallow water flow, we subsequently examine how the averaged field changes when the same finite window is applied directly to the fields or in the wave reference frame defined by the transformation \eqref{eq:wave-frame-variable}. Finally, we express Eulerian, Lagrangian and wave-frame averaging in a common formulation and compare the Eulerian and Lagrangian wave--mean decompositions with the exact projection to understand how the reference frame determines which motion is assigned to the mean.

\subsection{Modification of a classical Rossby wave example}\label{sec:andrews-example}

\begin{figure}[!tp]
  \centering
  \includegraphics[width=0.64\linewidth]{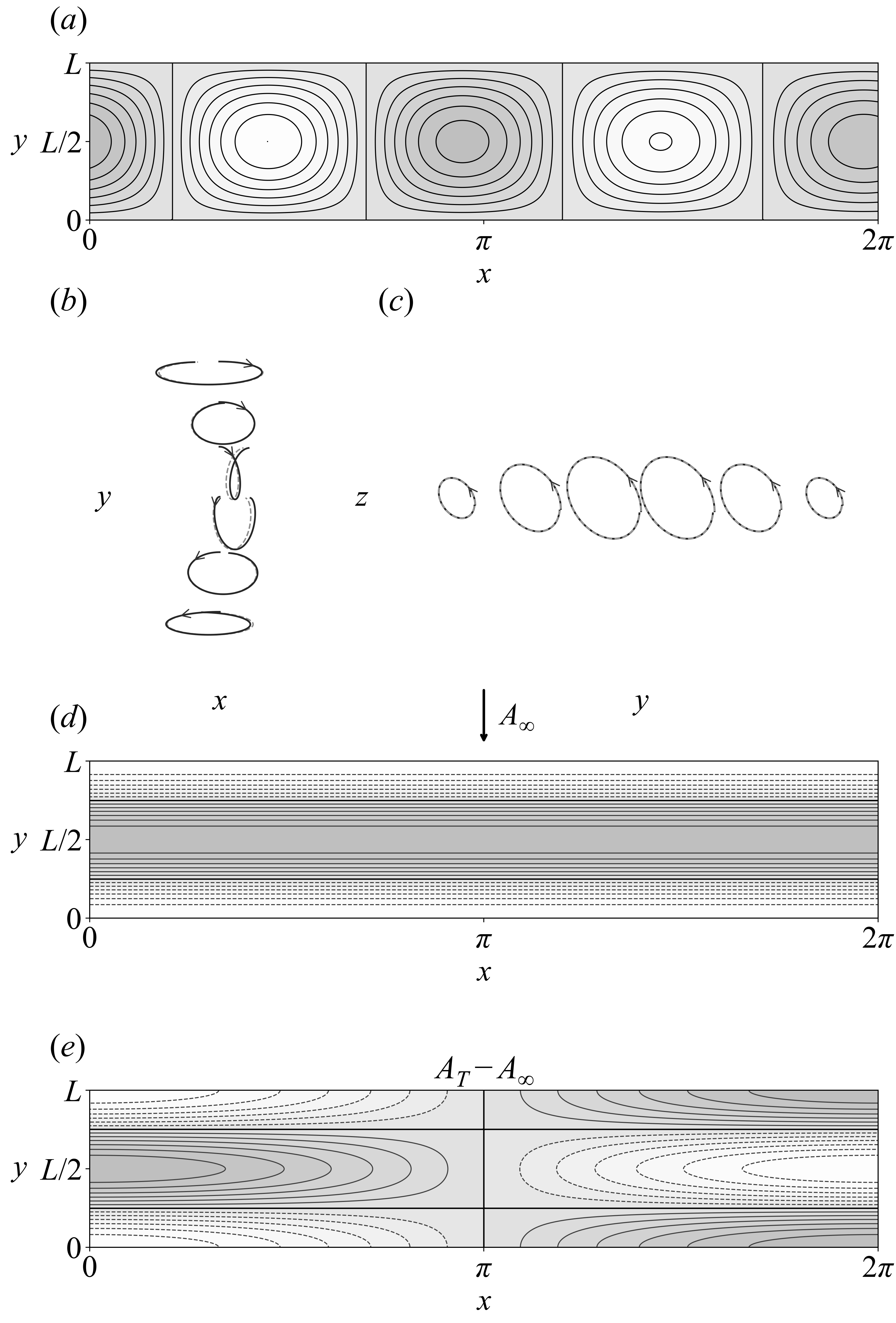}
  \caption{Two-frequency extension of the stationary Rossby-wave configuration in \citet[figure~4.13]{andrewsHoltonLeovy1987middleAtmDyn}, reproduced in figure~\ref{fig:andrews-single-wave}. Panel~(a) shows a snapshot of the resulting streamfunction \eqref{eq:modified-andrews-streamfunction} at $t=0$ and $z=0.2$, obtained by adding a second wave at a nearby zonal wavenumber. Because the drift is quadratic in wave displacement and velocity, it contains contributions at the difference frequency $\omega_2-\omega_1$ and at the higher frequencies $\omega_1+\omega_2$, $2\omega_1$, and $2\omega_2$. We isolate the contribution oscillating at $\omega_2-\omega_1$ as the nearly-slow drift and omit the higher-frequency contributions. Panels~(b) and (c) show reconstructed parcel orbits obtained by combining the first-order displacements of both waves with the horizontal drift integrated at fixed reference positions over one base-wave period; the solid curves use the exact drift together with the nearly-slow drift retained by the uniform time average, while the dashed curves use the exact drift alone. For visual clarity, displacements are enlarged by factors of $2$ in panel~(b) and $1.8$ in panel~(c), with a further fourfold zonal stretch in panel~(b); the drift-induced displacements are additionally enlarged by a factor of $1.71$. For panels~(b)--(e), we calculate the drift from the horizontal velocity at $z=0.2$. Panel~(c) shows the same line styles in the $y$--$z$ projection, with the vertical motion supplied by the first-order displacement of both waves; the small meridional separation is due entirely to the retained nearly-slow drift, since the exact drift is purely zonal. Panel~(d) shows the $x$-component of the exact drift, while panel~(e) shows the nearly-slow contribution to the zonal drift. The drift panels use a uniform time average over one period of the base wave, $T=2\pi/\omega_1$, with parameters chosen for illustration: $a_1=1$, $a_2=0.10$, $k_1=2$ and $k_2=2.5$. For the chosen window, the nearly-slow contribution at $\Delta\omega=\omega_2-\omega_1$ is weighted by $\operatorname{sinc}(\Delta\omega T/2)\approx0.972$.}
  \label{fig:andrews-modified}
\end{figure}

The single-wave stationary Rossby example recalled in figure~\ref{fig:andrews-single-wave} reproduces the configuration of \citet[figure~4.13]{andrewsHoltonLeovy1987middleAtmDyn}. For this example, we use $x$ for the zonal coordinate in a frame moving with the uniform background current $\bar U$. We use the definition of $P$ in \eqref{eq:exact-wave-filter}, with $\mathcal L_{\mathrm F}$ now taken to be the linear Rossby-wave operator in this frame. In this single-wave configuration, a uniform average over one wave period gives the same drift as the infinite-time average. Because the drift is quadratic in the single-wave fields, it contains only zero-frequency and $2\omega$ contributions, of which both averages retain only the former. Adding a second wave at a nearby frequency introduces a slowly varying drift contribution at $\Delta\omega$ that is only partly cancelled by the finite-time average but removed entirely by the infinite-time average. Keeping the original wave and its uniform background current unchanged, we add a weaker wave at a nearby zonal wavenumber and write the perturbation streamfunction $\psi'$ as
\begin{equation}\label{eq:modified-andrews-streamfunction}
\psi' = e^{z/(2H)}\sin(\ell y) \left( a_1\cos(k_1 x+mz+\omega_1 t) + a_2\cos(k_2 x+mz+\omega_2 t) \right),
\end{equation}
where $a_1$ is the amplitude of the original wave and $a_2$ is that of the added wave, with $k_2 = k_1 + \Delta k$. We determine both wave frequencies from the Rossby-wave dispersion relation:
\begin{equation}\label{eq:andrews-two-wave-dispersion}
\omega_j = \frac{\beta k_j}{k_j^2+\ell^2+\frac{f_0^2}{N^2}\left(m^2+\frac{1}{4H^2}\right)}, \qquad j=1,2,
\end{equation}
where $\beta$ is the meridional gradient of the Coriolis parameter, $f_0$ its reference value, $N$ the buoyancy frequency and $H$ the scale height. We retain $\beta$ from the single-wave example, so that $\omega_1=k_1\bar U$ and the original wave remains stationary in the laboratory frame. For the drift calculation, we use the horizontal wave velocity $\bm{u}_{\mathrm{fast}}=(-\partial_y\psi',\partial_x\psi')$ and obtain the horizontal displacement $\bm{\xi}_h$ as its time antiderivative at fixed $x$, with $A_\infty\bm{\xi}_h=0$. The drift contribution~\eqref{eq:drift-diagnostic} produced by their interaction oscillates at the difference frequency $\Delta\omega\defeq\omega_2-\omega_1$, which is small because the two frequencies are close. Because $\Delta\omega \neq 0$, the exact average removes the cross-frequency term entirely, whereas the finite-time average retains a window-dependent fraction. For a uniform time average, \eqref{eq:near-matching-sinc} gives this fraction as $\operatorname{sinc}(\Delta\omega T/2)$, which approaches unity as the oscillation period grows relative to the window length.

Figure~\ref{fig:andrews-modified} shows the nearly-slow drift arising from the retained interaction between the two waves at the frequency difference $\Delta\omega$, as described by~\eqref{eq:drift-residual}.
\begin{figure}[!tp]
  \centering
  \includegraphics[width=\linewidth]{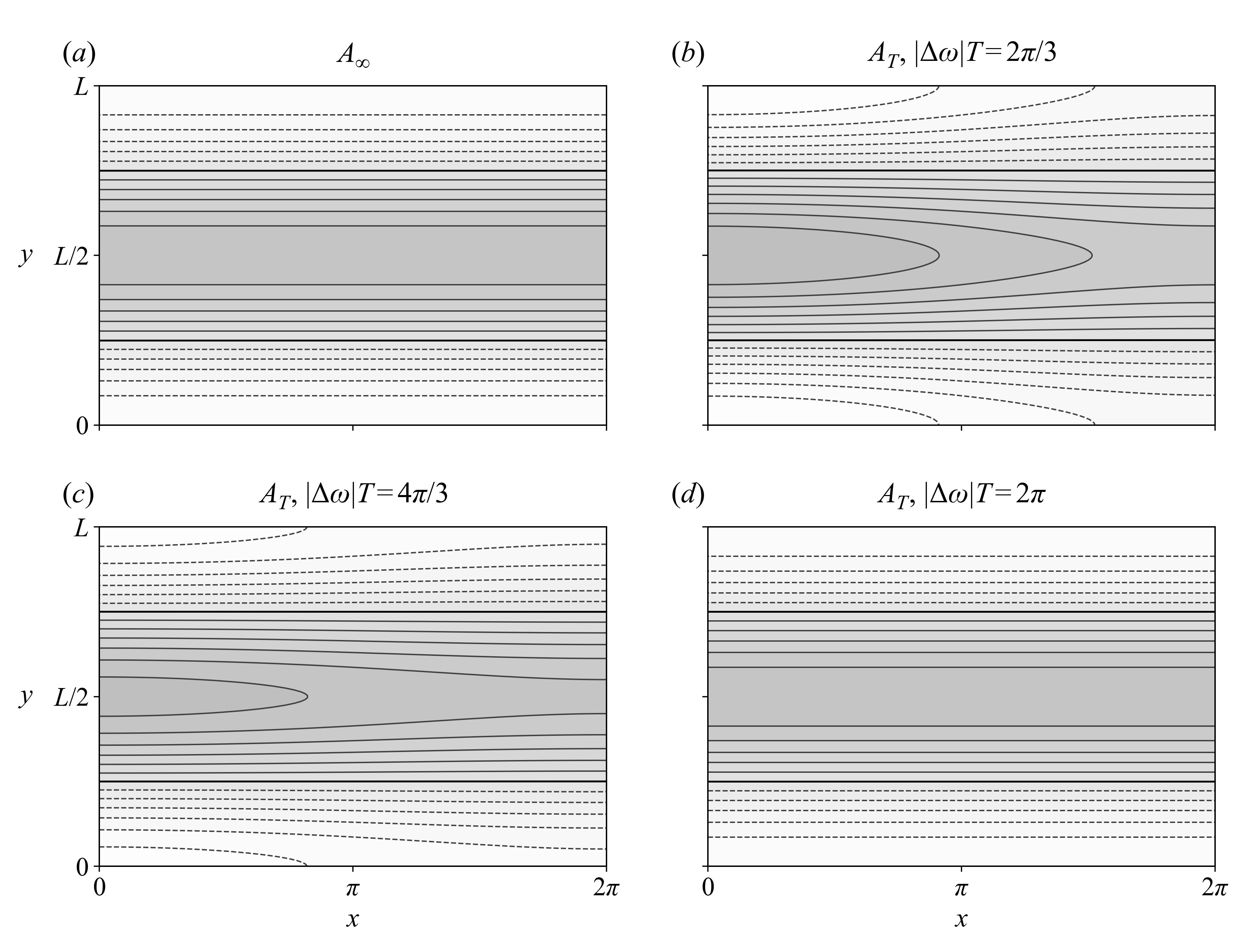}
  \caption{Window dependence of the drift for the two-frequency stationary Rossby-wave construction in figure~\ref{fig:andrews-modified}. Panel~(a) shows the zonal component $u^S_{A_\infty,x}$ of the exact drift $\bm{u}^S_{A_\infty}$. Panels~(b)--(d) show the exact zonal drift together with the portion of the nearly-slow cross-wave contribution retained by each averaging window; the rapidly oscillating sum- and double-frequency terms are omitted. We average uniformly over windows with $|\Delta\omega|T=2\pi/3,4\pi/3,2\pi$, using the same parameters as in figure~\ref{fig:andrews-modified}. Panel~(d) agrees with the exact drift to roundoff, because the window $|\Delta\omega|T=2\pi$ removes the contribution oscillating at $\Delta\omega$ under uniform averaging.}
  \label{fig:andrews-nonunique}
\end{figure} In panel~(a), the superposition of the two waves produces a slowly varying amplitude modulation on a scale set by $\Delta k$, much longer than the wavelength of either component. In panel~(b), the finite-window drift (here the exact drift plus the retained contribution at $\Delta\omega$) and the exact drift alone correspond to visibly different reconstructed parcel orbits. This contrasts with the single-wave case of figure~\ref{fig:andrews-single-wave}, where the two drifts agree and the loops are identical. The exact drift in panel~(d) varies only meridionally, because the exact average retains only the correlation of each wave with itself, not the cross-frequency correlation between the two. The nearly-slow drift in panel~(e) adds zonal structure that mirrors the amplitude modulation in panel~(a), taking the form of a coherent velocity field rather than a small residual. Moreover, the decomposition is not unique: because the cross-frequency weight depends on $T$, changing the window length can change the drift obtained from the same wave field.

This non-uniqueness is illustrated further by figure~\ref{fig:andrews-nonunique}, where the drifts over the two shorter windows (panels b, c) appear qualitatively similar, but over the longest window (panel d) the relative phase winds through one full cycle, $|\Delta\omega|T=2\pi$, causing the zonally varying part of the drift to cancel. Notably, choosing the averaging window also means choosing which transport contributions the diagnosed drift retains. The longest window removes the nearly-slow contribution and recovers the exact drift, whereas the two shorter windows retain it to different degrees. Because a broadband wave field contains many phase mismatches, no single finite window generally cancels them all; the drift estimated from a finite record can therefore retain a window-dependent nearly-slow contribution absent from the exact drift. Accordingly, the appropriate window length depends both on the transport one aims to retain and on how rapidly the slow state evolves, so no single value of $T$ is optimal for all purposes.

\subsection{Rotating shallow water equations}\label{sec:rsw-equations}

Having illustrated how the averaging window affects the drift, we now consider how the reference frame affects the averaged flow when the averaging window is held fixed. In a nonlinear rotating shallow water flow, we show that the same average gives different results depending on whether it is taken in the wave reference frame or applied to the fields in place.

The two-dimensional rotating shallow water equations for a single fluid layer of thickness $H+\eta$, with $H$ the mean depth and $\eta$ the free-surface displacement, are
\begin{subequations}\label{eq:rsw}
\begin{align}
\partial_t u + u\partial_x u + v\partial_y u - f v + g\partial_x \eta &= 0, \\
\partial_t v + u\partial_x v + v\partial_y v + f u + g\partial_y \eta &= 0, \\
\partial_t \eta + \partial_x\bigl((H+\eta)u\bigr) + \partial_y\bigl((H+\eta)v\bigr) &= 0.
\end{align}
\end{subequations}

Linearizing the rotating shallow water equations \eqref{eq:rsw} about rest gives a zero-frequency geostrophic mode and two inertia--gravity modes. For the state vector $\bm{w}_{\mathrm{RSW}}\defeq(u,v,\eta)^\top$ and horizontal wavevector $\bm{k}_H\defeq(k,l)$, the corresponding fast operator has the Fourier representation
\begin{equation}\label{eq:rsw-LF}
\mathcal{L}_{\mathrm F}(\bm{k}_H) = \begin{pmatrix} 0 & -f & i g k \\ f & 0 & i g l \\ i H k & i H l & 0 \end{pmatrix},
\end{equation}
with the fast frequency scale absorbed into $\mathcal{L}_{\mathrm F}$, so that $\partial_t\bm{w}_{\mathrm{RSW}}+\mathcal{L}_{\mathrm F}\bm{w}_{\mathrm{RSW}}=0$. The inertia--gravity (Poincar\'e) modes have frequencies $\pm\omega(\bm k_H)$, where $\omega(\bm k_H)=\sqrt{f^2+gH|\bm k_H|^2}$.

We nondimensionalize using the time scale $1/f$, horizontal length scale $\sqrt{gH}/f$, and thickness scale $H$, so that the nondimensional parameters satisfy $f=g=H=1$. For this experiment, we use a standard pseudospectral method on the doubly periodic domain $[0,2\pi]^2$. The equations \eqref{eq:rsw} are discretized on a uniform grid of $N^2=1024^2$ points, dealiased by the two-thirds rule, and advanced using the classical fourth-order Runge--Kutta method. We compare the lab-frame and wave-frame averages with the full field and its exactly projected balanced component. We also calculate the initial nonlinear forcing and use the weights derived in \S\ref{sec:exact-and-finite} to compare contributions from exactly resonant, near-resonant sideband and far-detuned interactions.

The flow is initialized with a steady balanced jet carrying three inertia--gravity waves:
\begin{equation}\label{eq:rsw-jet}
\bm{w}_{\mathrm{RSW}}(\bm{x},0) = \left(U\cos y,0,-\frac{f}{g}U\sin y\right)^\top+\operatorname{Re}\sum_{\bm{k}_H} a_{\bm{k}_H}\bm{r}_{\bm{k}_H}e^{i(\bm{k}_H\cdot\bm{x}+\phi_{\bm{k}_H})},
\end{equation}

with peak speed $U=0.08$ (chosen for illustration). The free-surface displacement of the jet was chosen so that the jet is in exact geostrophic balance, $fu=-g\partial_y\eta$.

The sum runs over $\bm{k}_H\in\{(4,0),(3,1),(3,-2)\}$, and $\bm{r}_{\bm{k}_H}$ is the positive-frequency Poincar\'e eigenvector of $\mathcal{L}_{\mathrm F}(\bm{k}_H)$, normalized so that its height component equals one. The phases and amplitudes are chosen to form a controlled narrowband packet with the $(4,0)$ wave dominant.

We compare two averages of this flow using the same scalar temporal weighting $K_T$ from \eqref{eq:kernel-average}: the first averages the fields directly, while the second averages in the wave reference frame before mapping the result back. We refer to the former method as the \emph{lab-frame average}
\begin{equation}\label{eq:lab-average}
\langle\bm{w}_{\mathrm{RSW}}\rangle_{\mathrm{lab}}(t;T) \defeq (A_T\bm{w}_{\mathrm{RSW}})(t) = \int K_T(s)\bm{w}_{\mathrm{RSW}}(t+s)\mathrm{d}s.
\end{equation}
We define the latter method of averaging as the \emph{wave-frame average}
\begin{equation}\label{eq:wave-average}
\langle\bm{w}_{\mathrm{RSW}}\rangle_{\mathrm{wave}}(t;T) \defeq e^{-t\mathcal{L}_{\mathrm F}}\int K_T(s)e^{(t+s)\mathcal{L}_{\mathrm F}}\bm{w}_{\mathrm{RSW}}(t+s)\mathrm{d}s,
\end{equation}
which maps each instant into the wave reference frame, averages there, and maps back. The wave-frame average can be written in two additional equivalent forms, given that $\mathcal L_{\mathrm F}$ is time independent.
\begin{subequations}\label{eq:wave-average-kernel-forms}
First, we incorporate the inner mapping into an operator-valued kernel,
\begin{equation}\label{eq:wave-average-inner-kernel}
\langle\bm{w}_{\mathrm{RSW}}\rangle_{\mathrm{wave}}(t;T)=e^{-t\mathcal L_{\mathrm F}}\int \widetilde K_T(t,s)\bm{w}_{\mathrm{RSW}}(t+s)\mathrm{d}s,
\end{equation}
where $\widetilde K_T(t,s)\defeq K_T(s)e^{(t+s)\mathcal L_{\mathrm F}}$. Second, both the inner and outer mappings may be considered as part of an operator-valued kernel, so that
\begin{equation}\label{eq:wave-average-combined-kernel}
\langle\bm{w}_{\mathrm{RSW}}\rangle_{\mathrm{wave}}(t;T)=\int K_T^{\mathrm W}(s)\bm{w}_{\mathrm{RSW}}(t+s)\mathrm{d}s,
\end{equation}
where $K_T^{\mathrm W}(s)\defeq K_T(s)e^{s\mathcal L_{\mathrm F}}$.
\end{subequations}

Because a freely propagating wave satisfies $\partial_t\bm{w}_{\mathrm{RSW}}+\mathcal{L}_{\mathrm F}\bm{w}_{\mathrm{RSW}}=0$, samples separated by $s$ carry different fast phases and are related by $\bm{w}_{\mathrm{RSW}}(t+s)=e^{-s\mathcal{L}_{\mathrm F}}\bm{w}_{\mathrm{RSW}}(t)$. Because of these phase differences, the wave oscillations partially cancel in the lab-frame average. In this free-wave limit, the wave-frame average instead maps every sample to the same state $\bm{w}_{\mathrm{RSW}}(0)$ before averaging and subsequently restores the wave's phase at time $t$, leaving the wave unchanged. Under the linear fast dynamics, the infinite-window limit of the lab-frame average recovers the balanced component selected by the exact projection $P$ in \eqref{eq:exact-wave-filter}. In the nonlinear flow, the amplitudes continue to evolve in the wave reference frame, so averaging smooths their modulation before the fast phase is restored.

\begin{figure}[!t]
  \centering
  \includegraphics[width=0.90\linewidth]{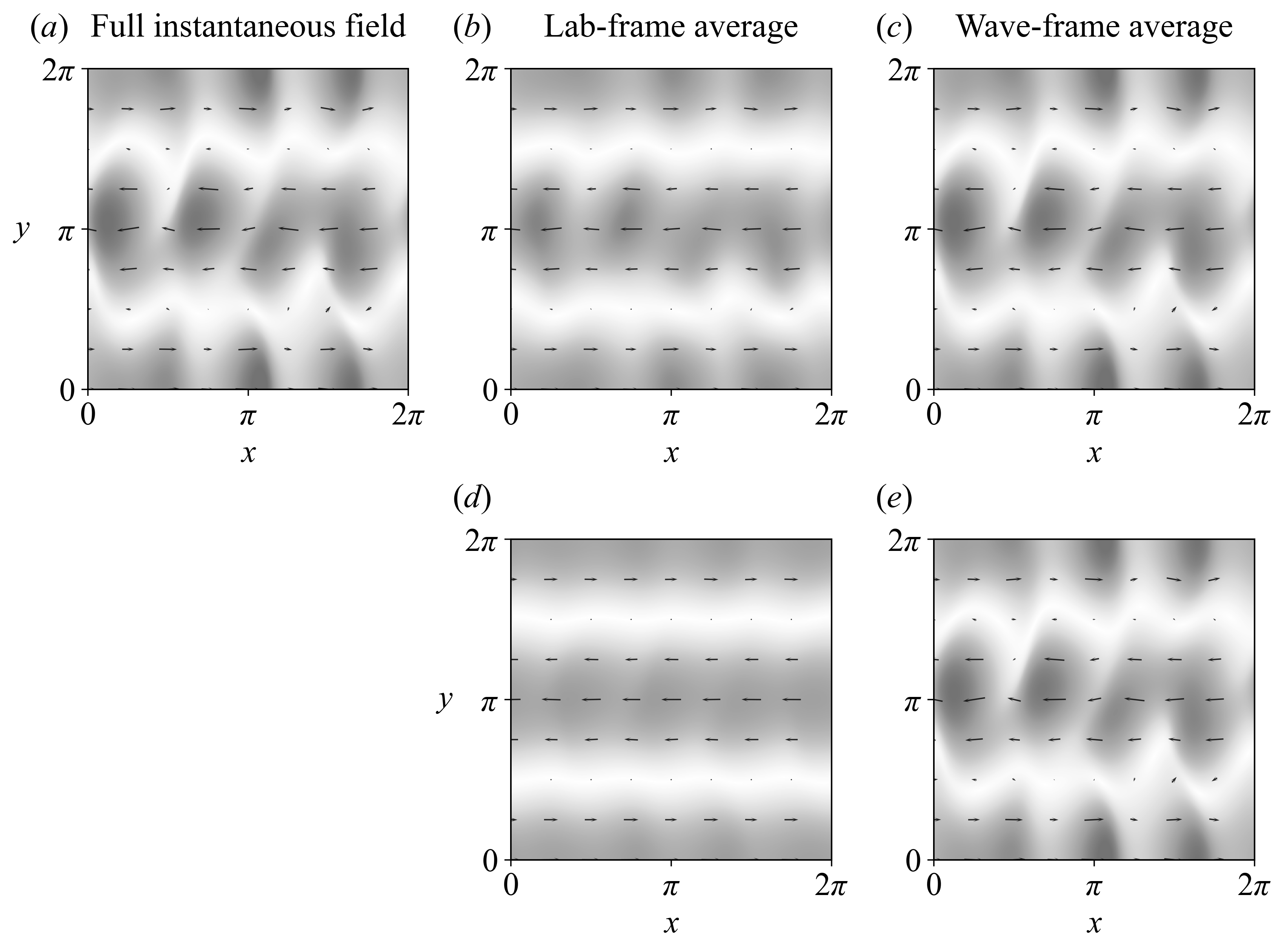}
  \caption{Velocity field (arrows) over grayscale speed shading for the rotating shallow water example at $t=3$, formed using a centered uniform time average (boxcar kernel) with window lengths $T=1$ (top row) and $T=2$ (bottom row) --- about $0.6$ and $1.2$ wave periods: ($a$)~the full instantaneous field, ($b$,$d$)~the lab-frame average, and ($c$,$e$)~the wave-frame average. Averaging the fields in place attenuates the wave pattern, the more so the longer the window ($b$,$d$), whereas the wave-frame average preserves it at both windows ($c$,$e$).}
  \label{fig:rsw-velocity-fields}
\end{figure}

\begin{figure}[!htbp]
  \centering
  \includegraphics[width=0.82\linewidth]{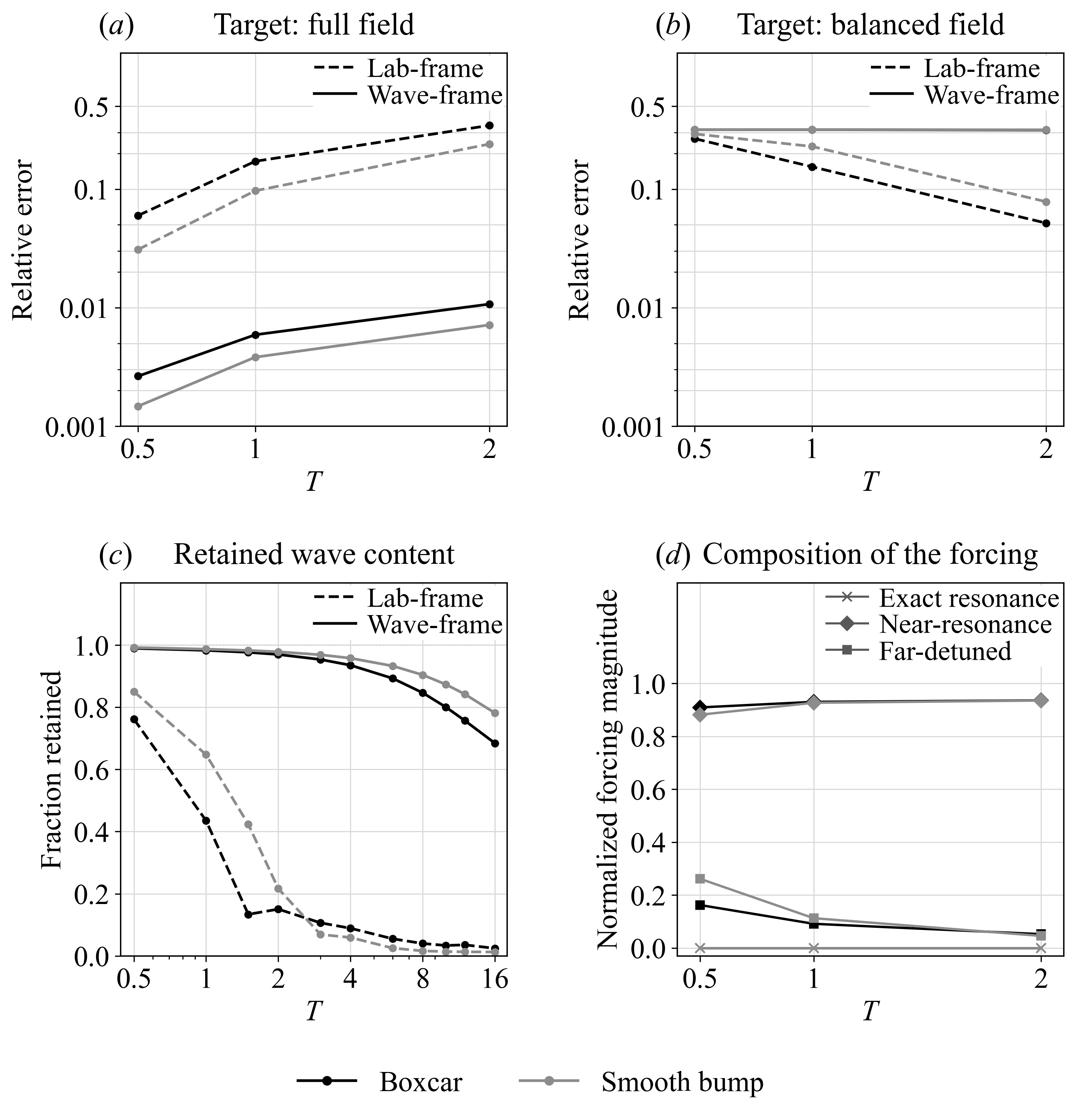}
  \caption{A comparison of the wave-frame and lab-frame averages in figure~\ref{fig:rsw-velocity-fields} through their relative errors and retained wave content. ($a$)~Relative error against the full instantaneous field; ($b$)~relative error against the slow balanced field; ($c$)~the fraction of the full field's wave content retained by each average. All errors and retained-wave fractions use the domain root-mean-square Euclidean norm of the nondimensional state $(u,v,\eta)^\top$: each error is the norm of the difference from the reference field divided by the reference-field norm, while each retained-wave fraction is the ageostrophic norm of the averaged field divided by that of the full field. In ($a$--$c$), solid and dashed lines denote the wave-frame and lab-frame averages, with black and gray indicating the boxcar and smooth-bump kernels. Panels ($a$,$b$,$d$) show $T\le2$, with averages centered at $t=3$ in ($a$,$b$) and triad contributions evaluated at the initial state in ($d$). Panel ($c$) extends the range of $T$, with averages centered at $t=8$. For each triad group, we sum its finite-window-weighted contributions to each mode's evolution and normalize the resulting vector's Euclidean norm by that obtained from all triads. In ($d$), exactly resonant, near-resonant sideband, and far-detuned contributions are marked by crosses, diamonds, and squares, respectively. The steady jet scatters the primary mode $\bm{k}_H=(4,0)$ into $(4,\pm1)$, producing the near-resonant sideband at $|\Delta\omega|\approx0.12$. For this comparison, we classify interactions with $|\Delta\omega|\ge4/T$ (the factor of four being arbitrary) as far-detuned. The exactly resonant contribution is nonzero but negligible on this scale, because those triads couple only modes the initial packet leaves essentially empty. As the window lengthens, the wave-frame average stays close to the full field and retains most of its wave content, whereas the lab-frame average approaches the balanced field and retains progressively less.}
  \label{fig:rsw-dns-observables}
\end{figure}

We compare each average \eqref{eq:lab-average} and \eqref{eq:wave-average} with the full instantaneous state $\bm{w}_{\mathrm{RSW}}$ from which it is formed and with the balanced field. The balanced field is again the slow flow that remains once the inertia--gravity oscillations are removed, calculated via
\begin{equation}
\bm{w}_{\mathrm{bal}}(t) \defeq P \bm{w}_{\mathrm{RSW}}(t).
\end{equation}
For each average, the retained wave motion is measured via the root-mean-square ageostrophic part of the field relative to that of the full field.

Figure~\ref{fig:rsw-velocity-fields} compares the full and averaged velocity fields for window lengths $T=1$ and $T=2$. Direct averaging tends to smear out the wave field, whereas averaging in the wave reference frame retains the coherent wave pattern and smooths its slower modulation. By $T=2$, about $1.2$ wave periods, the lab-frame field has lost nearly all its wave structure, while the wave-frame field still carries the full pattern. The wave-frame average tracks the full field closely but is not equal to it. The two coincide only for freely propagating waves, whereas here interactions slowly modify the wave envelopes.

To quantify these differences, figure~\ref{fig:rsw-dns-observables} compares each average with the full state and its balanced component. The weighting across a finite window controls how strongly the wave oscillations cancel. We compare a time average that gives equal weight to all instants within the window (boxcar kernel) with one whose weight tapers smoothly to zero at the window endpoints (smooth bump kernel, $K(s)\propto e^{-1/(1-4s^2)}$ for $|s|<1/2$, zero otherwise). Relative to the full field, the wave-frame error remains within a few percent, while the lab-frame error reaches tens of percent (figure~\ref{fig:rsw-dns-observables}$a$). The wave-frame error grows with $T$ just as averaging over a longer window often increases the departure from a single instantaneous field; even so, it stays several times below the lab-frame error for every window. The lab-frame error remains larger because the window is applied before the fast phase is removed, partially canceling the oscillatory component rather than smoothing only its slowly varying envelope. As the window length increases, the lab-frame average approaches the balanced field, whereas the wave-frame average stays at nearly constant relative error (figure~\ref{fig:rsw-dns-observables}$b$).

The wave content retained by the lab-frame average depends on the kernel's frequency response $\widehat{K}(T\omega)$, as given by \eqref{eq:spectral-multiplier}. At $T=2$, the boxcar and smooth-bump lab-frame averages retain $15\%$ and $22\%$ of the wave content, respectively, while the corresponding wave-frame averages retain $97\%$ and $98\%$ (figure~\ref{fig:rsw-dns-observables}$c$). Across both kernels, the dominant difference is therefore the averaging frame rather than the temporal weighting within the window. Figure~\ref{fig:rsw-dns-observables}$d$ shows that the finite-window forcing is dominated by a near-resonant sideband, produced when the balanced jet scatters the primary wave into neighbouring modes. Thus, in this flow, infinite-time averaging removes the near-resonant interactions that contribute most strongly to the amplitudes' nonlinear rate of change at the initial state.

Over these windows, the lab-frame average \eqref{eq:lab-average} is closest to the balanced component, while the wave-frame average \eqref{eq:wave-average} is closest to the full instantaneous state. An important implication of this comparison is that no single reference frame is optimal for every averaging task, since the choice depends on which motions the mean is intended to represent.

\subsection{Wave--mean decomposition across reference frames}\label{sec:reference-frame-comparison}

In a final numerical comparison, we place the decomposition associated with the wave reference frame, defined by the exact projection $P$, alongside decompositions obtained by averaging in Eulerian and Lagrangian reference frames. Here we ask whether differences among these decompositions can be traced to which structures are slowly varying in each frame.

Recent work has developed Lagrangian filtering to address limitations of Eulerian averaging in wave--mean decomposition \citep[e.g.][]{nagai2015,shakespearehogg2017,shakespeare2021,kafiabadvanneste2023,bakerkafiabadmaitlanddaviesvanneste2025}. To construct a wave--mean decomposition using Lagrangian filtering, \citet{bakerkafiabadmaitlanddaviesvanneste2025} introduce a mean obtained by averaging a scalar field along material trajectories; writing their filter (equation (3.2) therein) in the convention of \eqref{eq:kernel-average} yields
\begin{equation}\label{eq:lagrangian-midpoint-trajectory}
(A_T^{\mathrm L}\theta)(\bm{x},t)\defeq\int_{\mathbb R}K_T(s)\theta\bigl(\bm{X}(s;\bm{x},t),t+s\bigr)\mathrm{d}s.
\end{equation}
Here $\theta$ is the scalar field being filtered (or decomposed), $\bm{X}(s;\bm{x},t)$ is the position at $t+s$ of the particle located at $\bm{x}$ at $t$, and $K_T$ is a normalized finite-window kernel. The superscript $\mathrm L$ is used to distinguish this average as the representative Lagrangian average. The reference time $t$ is the midpoint of the averaging interval, so $\bm{x}$ labels a particle's (or parcel's) position at that midpoint. Note that this convention differs from \S\ref{sec:drift}, where $A_\infty\bm\xi=0$; using the midpoint position assigns each particle's trajectory average to its position in the instantaneous snapshot. The resulting Lagrangian mean can therefore retain wave-induced displacements of mean structures. All three methods then decompose that snapshot in the same spatial coordinates, allowing a direct comparison of which structures they assign to the mean and waves.

The material derivative $D_{\mathrm L}\defeq\partial_t+\bm{u}\cdot\nabla$ generates the evolution of $\theta$ along $\bm{X}$ (Appendix~\ref{app:averaging-generator-identities}), so that
\begin{equation}\label{eq:material-characteristic-evolution}
\bigl(e^{sD_{\mathrm L}}\theta\bigr)(\bm{x},t)=\theta\bigl(\bm{X}(s;\bm{x},t),t+s\bigr).
\end{equation}
Using \eqref{eq:material-characteristic-evolution}, the Lagrangian average \eqref{eq:lagrangian-midpoint-trajectory} may be written directly in terms of the material evolution:
\begin{equation}\label{eq:lagrangian-midpoint-generator}
(A_T^{\mathrm L}\theta)(\bm{x},t)=\int_{\mathbb{R}}K_T(s)\bigl(e^{sD_{\mathrm L}}\theta\bigr)(\bm{x},t)\mathrm{d}s.
\end{equation}

An Eulerian average samples the scalar field $\theta$ at a fixed spatial position, so its reference evolution is ordinary time translation. With $D_{\mathrm E}\defeq\partial_t$, we define
\begin{equation}\label{eq:eulerian-average}
(A_T^{\mathrm E}\theta)(\bm{x},t)\defeq\int_{\mathbb R}K_T(s)\theta(\bm{x},t+s)\mathrm{d}s=\int_{\mathbb R}K_T(s)\bigl(e^{sD_{\mathrm E}}\theta\bigr)(\bm{x},t)\mathrm{d}s.
\end{equation}

To compare finite-window averaging across the three reference frames, we use the operator-valued kernel form \eqref{eq:wave-average-combined-kernel} of the wave-frame average. Defining $D_{\mathrm W}\defeq\partial_t+\mathcal L_{\mathrm F}$, we denote the same wave-frame average of a generic state $\bm{w}$ by
\begin{equation}\label{eq:wave-average-operator-form}
(A_T^{\mathrm W}\bm{w})(t)\defeq\int_{\mathbb R}K_T^{\mathrm W}(s)\bm{w}(t+s)\mathrm{d}s=\int_{\mathbb R}K_T(s)e^{sD_{\mathrm W}}\bm{w}(t)\mathrm{d}s.
\end{equation}

The Eulerian, Lagrangian, and wave-frame averages may now be easily compared:
\begin{subequations}\label{eq:frame-generator-filters}
\begin{align}
A_T^{\mathrm E}\theta&=\int_{\mathbb R}K_T(s)e^{sD_{\mathrm E}}\theta\,\mathrm{d}s,\label{eq:eulerian-generator-filter}\\
A_T^{\mathrm L}\theta&=\int_{\mathbb R}K_T(s)e^{sD_{\mathrm L}}\theta\,\mathrm{d}s,\label{eq:lagrangian-generator-filter}\\
A_T^{\mathrm W}\bm{w}&=\int_{\mathbb R}K_T(s)e^{sD_{\mathrm W}}\bm{w}\,\mathrm{d}s.\label{eq:wave-generator-filter}
\end{align}
\end{subequations}
The Eulerian and Lagrangian averages are applied to the scalar $\theta$ following \citet{bakerkafiabadmaitlanddaviesvanneste2025}, while the wave-frame average is applied to the vector $\bm{w}$, on which $\mathcal L_{\mathrm F}$ acts, consistent with our treatment throughout the paper. Note that, as in \eqref{eq:wave-average-combined-kernel}, the scalar temporal weighting may be combined with each reference-frame transformation into an operator-valued kernel. Writing the three filters in the form \eqref{eq:frame-generator-filters} highlights their common construction, in which samples from different times are aligned through the chosen reference-frame transformation and averaged with the same scalar temporal weighting $K_T$. Just as one may average at a fixed spatial position or while following a particle along its trajectory, one may average while following the waves under the fast linear evolution.

Because we consider Eulerian and Lagrangian filtering specifically as methods for wave--mean decomposition in the comparison that follows, we compare them with the exact projection $P$, rather than with the finite wave-frame average, which retains the wave signature (figures~\ref{fig:rsw-velocity-fields} and \ref{fig:rsw-dns-observables}). Operationally, the finite wave-frame average acts on an evolving history after its fast phase has been removed, whereas $P$ averages the inverse fast-wave orbit of an instantaneous state over infinite time, as in \eqref{eq:exact-wave-filter}.

\begin{figure}[!t]
  \centering
  \includegraphics[width=\linewidth]{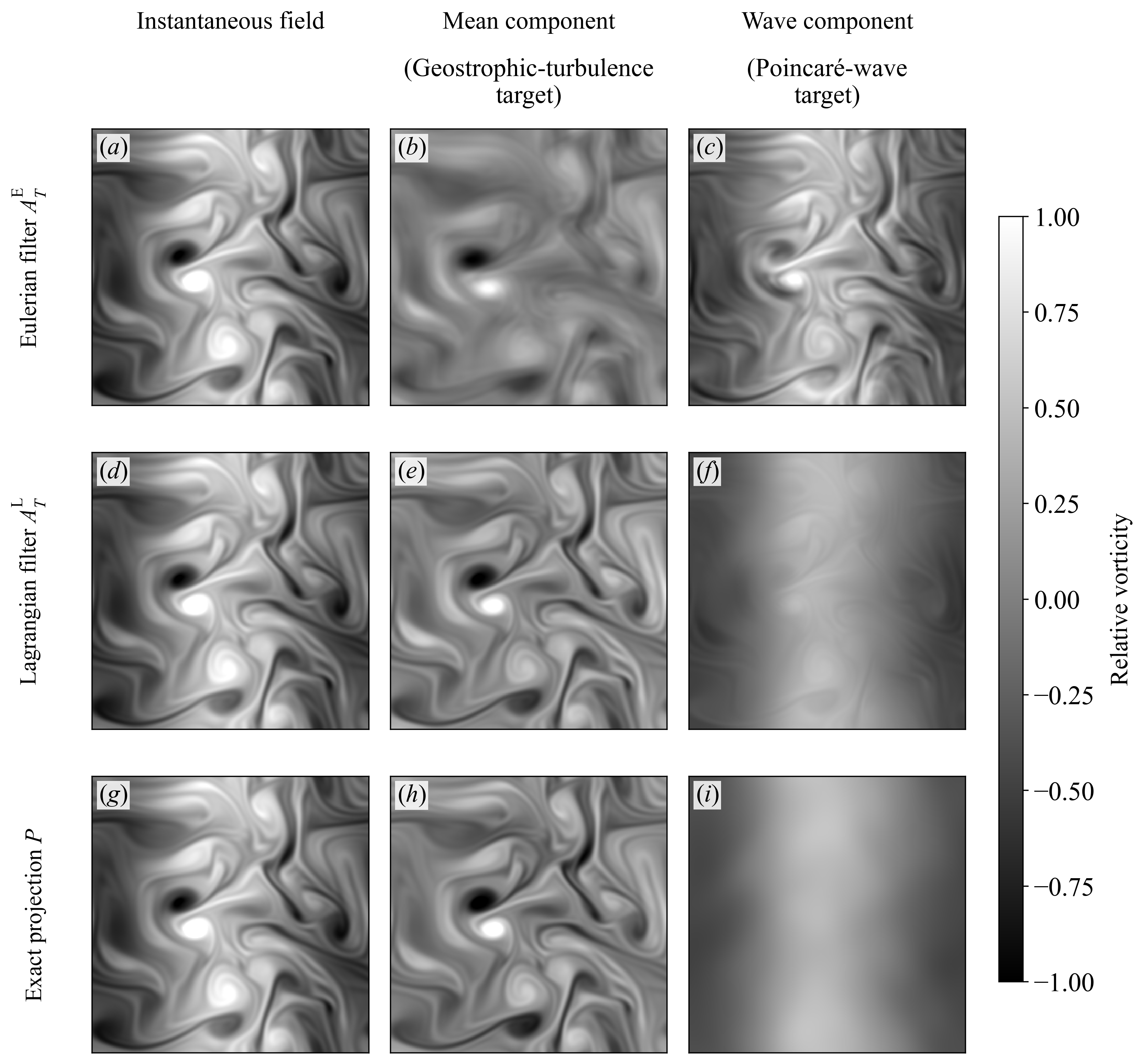}
  \caption{Wave--mean decompositions of relative vorticity in the modified shallow water simulation of \citet{bakerkafiabadmaitlanddaviesvanneste2025}, evaluated at $t=20$, the midpoint of the averaging window of length $T=40$. Panels ($a$--$c$), ($d$--$f$), and ($g$--$i$) show the decompositions produced by the Eulerian filter $A_T^{\mathrm E}$, the Lagrangian filter $A_T^{\mathrm L}$, and the exact projection $P$, respectively. Panels ($a$--$c$) and ($d$--$f$) replot the corresponding Eulerian and Lagrangian decompositions from their figure~5($a$--$c$,$g$--$i$) using the data archived with that study; panels ($g$--$i$) show the corresponding decomposition obtained in the present study with the exact projection $P$. Within each row, the panels show the common instantaneous field, its mean component, and its wave component from left to right. The Eulerian and Lagrangian rows apply the same low-pass filter with cutoff frequency $2$, truncated to this interval and renormalized to unit integral, to relative-vorticity records at fixed positions and along material trajectories, respectively, whereas the projection row applies $P$ jointly to the velocity and free-surface displacement, $(u,v,\eta)^\top$, at the temporal midpoint before diagnosing relative vorticity from the mean and wave velocities. Note that obtaining the projected decomposition is computationally practical here: in matched $N=256$ runs, evolving to the temporal midpoint and applying $P$ required $325\,\mathrm{s}$ of wall time, compared with $1505\,\mathrm{s}$ for the full-window Lagrangian calculation. The improved separation under Lagrangian filtering in this example is consistent with its closer approximation to $P$: relative to panel ($i$), the wave fields in panels ($f$) and ($c$) have Pearson correlations $0.993$ and $0.888$, and relative $L^2$ differences $0.191$ and $0.519$, respectively.}
  \label{fig:reference-frame-decompositions}
\end{figure}

To understand why the decompositions differ, we compare how rapidly the same physical motion varies in each reference frame. Because a motion that is slow in one frame may be rapid in another, it can enter the mean component in one decomposition and the wave component in another. The sampled field, evaluated at times $t+s$ with $s$ measured from the window centre $t$, is $\theta(\bm{x},t+s)$ at a fixed spatial position and $\theta(\bm{X}(s;\bm{x},t),t+s)$ along a material trajectory. Taking the temporal Fourier transform of each record with respect to $s$ gives its frequency spectrum. We use $\omega_{\mathrm E}$ and $\omega_{\mathrm L}$ for temporal frequencies in the two records, corresponding to sampling at a fixed position and along a material trajectory, respectively. Each record can contain many frequencies, with its own amplitude and phase at each frequency. Each mean is obtained by combining the Fourier contributions from its time record, weighted by the response of the finite-window kernel $K_T$ at $\omega_{\mathrm E}$ in the Eulerian case and at $\omega_{\mathrm L}$ in the Lagrangian case (see~\eqref{eq:app-frame-frequency-response}). Because $K_T$ is normalized, both averages preserve zero-frequency components, but they can assign different weights to the same physical motion since its frequency can differ between the two records.

To identify the part of the instantaneous state that is invariant under the fast linear evolution, we expand the state in eigenmodes of $\mathcal L_{\mathrm F}$. Each mode has an associated eigenfrequency $\omega_{\mathrm F}$, which we call its intrinsic frequency because it describes oscillation under the selected fast linear dynamics alone. The projection $P$ retains precisely those modes for which $\omega_{\mathrm F}=0$.

While each method centres its mean selection on zero frequency, the physical meaning of zero differs among the three evolutions. The frequency $\omega_{\mathrm L}$ measures how a component varies as it is sampled along a material trajectory; zero frequency therefore means that its value remains constant as the particle moves. The frequency $\omega_{\mathrm E}$ measures how a component varies as it is sampled at a fixed spatial position; a component with $\omega_{\mathrm E}=0$ is therefore steady at that position. The intrinsic frequency $\omega_{\mathrm F}$ measures how an eigenmode oscillates under the selected fast linear evolution; modes with $\omega_{\mathrm F}=0$ are invariant under the fast linear wave evolution.

To illustrate why Lagrangian averaging can retain balanced motion in agreement with $P$, while Eulerian averaging can attenuate it, we can consider a simple example of a single Fourier eigenmode of the shallow water fast dynamics \eqref{eq:rsw-LF} carried by a uniform flow at velocity $\bm U$. In the Eulerian frame, we write the mode as $\bm w(\bm x,t)=\widehat{\bm w}_{\bm k_H}e^{i(\bm k_H\cdot\bm x-\omega_{\mathrm E}t)}$. The mode satisfies $(\partial_t+\bm U\cdot\nabla+\mathcal L_{\mathrm F})\bm w=0$, with $\mathcal L_{\mathrm F}(\bm k_H)\widehat{\bm w}_{\bm k_H}=i\omega_{\mathrm F}\widehat{\bm w}_{\bm k_H}$. Substituting the plane-wave form into the governing equation gives $(-i\omega_{\mathrm E}+i\bm k_H\cdot\bm U+i\omega_{\mathrm F})\widehat{\bm w}_{\bm k_H}=0$. For a mode with nonzero amplitude, this gives $\omega_{\mathrm E}=\bm k_H\cdot\bm U+\omega_{\mathrm F}$. To obtain the Lagrangian frequency, we follow the material trajectory, which at linear order is $\bm X(s;\bm x,t)=\bm x+\bm U s$. Evaluating the mode at this position and time $t+s$ gives $\bm w(\bm X(s;\bm x,t),t+s)=\bm w(\bm x,t)e^{-i(\omega_{\mathrm E}-\bm k_H\cdot\bm U)s}$, whose frequency is $\omega_{\mathrm L}=\omega_{\mathrm E}-\bm k_H\cdot\bm U$. Combining these relations gives
\begin{equation}\label{eq:uniform-flow-frame-signals}
\omega_{\mathrm L}=\omega_{\mathrm E}-\bm k_H\cdot\bm U=\omega_{\mathrm F}.
\end{equation}
A balanced component carried by the uniform flow has $\omega_{\mathrm L}=\omega_{\mathrm F}=0$, even though a fixed observer measures $\omega_{\mathrm E}=\bm k_H\cdot\bm U$; a Poincar\'e wave carried by the same flow has $\omega_{\mathrm L}=\omega_{\mathrm F}=\pm\omega(\bm k_H)$, where $\omega(\bm k_H)=\bigl(f^2+gH|\bm k_H|^2\bigr)^{1/2}$ is the Poincar\'e-frequency magnitude, while its Eulerian frequency is Doppler shifted to $\bm k_H\cdot\bm U\pm\omega(\bm k_H)$. The balanced component therefore receives unit weight from both Lagrangian averaging and $P$, whereas it varies at frequency $\bm k_H\cdot\bm U$ as it is advected past a fixed point and can consequently be attenuated by $K_T$ under Eulerian averaging. Because $P$ selects only modes with zero intrinsic frequency, it removes the Poincar\'e wave, while Lagrangian and Eulerian averaging apply the response of $K_T$ at the intrinsic frequency $\omega_{\mathrm F}$ and the Doppler-shifted frequency $\bm k_H\cdot\bm U+\omega_{\mathrm F}$, respectively; a nonzero response at either frequency leaves wave content in the corresponding mean. Although idealized, this example suggests that, if Lagrangian averaging gives a wave--mean decomposition that better matches the expected separation between balanced motion and fast waves than does Eulerian averaging, it may be because sampling along particle paths removes the apparent time dependence of structures swept past a fixed point, leaving a distinction between intrinsic wave motion and mean flow that more closely matches the exact projection.

We now compare the Eulerian, Lagrangian, and projection-based decompositions directly using the modified shallow water simulation of \citet{bakerkafiabadmaitlanddaviesvanneste2025}. This model is well suited to the comparison because it prevents nonlinear wave steepening from producing shocks that would complicate the separation of turbulence (the mean flow in this example) and waves, while preserving the linear dynamics used to define $P$. The simulation is initialized with fully developed two-dimensional turbulence in geostrophic balance, together with a superimposed Poincar\'e wave with horizontal wavevector $\bm k_H=(1,0)$ \citep{bakerkafiabadmaitlanddaviesvanneste2025}. The balanced turbulence is slow in time but spans a range of spatial scales, including the fine vorticity filaments visible in figure~\ref{fig:reference-frame-decompositions}. Conversely, the Poincar\'e wave oscillates rapidly despite being domain scale. Thus, in these relative-vorticity fields, the visually intricate structure is associated with the mean target and the smooth large-scale pattern with the wave target. Because the Eulerian and Lagrangian filters do not use the known shallow water dispersion relation, we likewise construct $P$ directly from the discretized fast operator, keeping the comparison independent of analytic information that may be unavailable in practical applications. In this comparison, the Eulerian and Lagrangian filters are applied directly to relative vorticity, whereas $P$ acts jointly on the velocity and free-surface displacement, $(u,v,\eta)^\top$, at the temporal midpoint of the averaging window before relative vorticity is diagnosed from the mean and wave velocity fields.

In figure~\ref{fig:reference-frame-decompositions}, the Lagrangian decomposition agrees much more closely with the projected decomposition than does the Eulerian decomposition. The Lagrangian and projected wave components are dominated by the large-scale Poincar\'e-wave pattern, while their mean components preserve the filamentary turbulent structure. Eulerian filtering instead leaves a strong turbulent imprint in the wave component and blurs that structure in the mean. Thus, in this example, filtering along material trajectories over a finite window yields a spatial decomposition close to that obtained by projecting an instantaneous state onto the modes with zero frequency under the fast linear evolution. A significant benefit of expressing these averages in a common form is that their differences can be interpreted through the motions that vary slowly in each reference frame.

\section{Discussion}\label{sec:discussion}

A central difficulty in wave--mean decomposition is that one generally cannot fully separate all `wave' from all `mean' dynamics throughout their evolution in the nonlinear flows of nature. Even the average over all wave periods, the exact projection of \eqref{eq:exact-wave-filter}, gives an exact separation of wave content from mean content in the fields, but only a leading-order separation in the dynamics \citep{embidmajda1996,embidmajda1998}. At higher order, however, the waves enter the evolution of the slow remainder through exact and near-resonant triads and the pathways they compose.

In practice, averaging is often restricted to a finite interval, and here we examine how this departure from the classical infinite-time average retains near-resonant content that the exact separation removes. In \S\ref{sec:governing-fast-structure}, we express this exact average in terms of the fast linear dynamics that also define the wave reference frame. We generalize this construction to finite windows in \S\ref{sec:exact-and-finite}, broadening exact resonance selection and weighting the retained interactions by the frequency response of the averaging kernel. We illustrate the physical implications of this generalization through wave-induced drift (\S\ref{sec:drift}), showing how the wave interactions retained by the average contribute to transport. Expressing the drift as a sum over wave-mode pairs recovers the classical Stokes drift under exact averaging and identifies an additional nearly-slow contribution under finite averaging. The example of \S\ref{sec:andrews-example} illustrates the effect of including the nearly-slow drift by comparing parcel orbits with and without this contribution. Averaging in place can produce a field closer to the balanced flow, whereas mapping into the wave reference frame, averaging there and mapping back can better retain the coherent wave motion (\S\ref{sec:rsw-equations}). We express Eulerian, Lagrangian and wave-frame averaging in a common formulation (\S\ref{sec:reference-frame-comparison}) to show how the reference frame determines which motions vary slowly and therefore enter the mean. Together, these results show that the flow structures and wave-induced effects retained by a finite-time average depend on both the reference frame and the length and shape of the averaging window.

\subsection{Averaging and reference frames}\label{sec:discussion-averaging-frame}

The viewpoint developed here for wave--mean decomposition is that any such separation depends on whether an average can distinguish some oscillations from others over the time available, while the reference frame determines what it physically means for an oscillation to be slow. Intuitively, a mean flow is often considered as such because it persists in space and time. An underlying balance of forces can make this persistence possible. In a balanced asymptotic expansion, the mean is the leading-order solution obtained from this force balance. Waves are the restoring motions associated with departures from this balance. Averaging to separate the balanced mean from the restoring wave motions is therefore tied to the underlying force balance. The force balance is itself tied to a reference frame, and therefore so is the averaging. The result of \S\ref{sec:governing-fast-structure} makes this connection explicit because the same fast generator defines both the projection onto the balanced mean and the coordinate change defining the wave reference frame, although the two constructions use the fast evolution differently. Furthermore, this shared dependence on the fast generator explains why the choice of reference frame affects how well an average isolates the motions of interest. For instance, in \S\ref{sec:rsw-equations} we find that averaging in place produces a field closer to the balanced mean, whereas mapping into the wave reference frame, averaging there and mapping back retains the coherent wave motion. In the broader context of reference frames, we find closer agreement with the exact projection for the decomposition defined relative to the Lagrangian reference frame than for that defined relative to the Eulerian reference frame in the example of \S\ref{sec:reference-frame-comparison}. This suggests that the advantages observed here associated with a Lagrangian wave--mean decomposition may reflect the extent to which the separation induced by the Lagrangian reference frame matches the exact projection associated with the wave reference frame.

Distinguishing waves from the mean can be understood as a frequency question: waves are fast relative to the mean, the mean is near zero frequency, and a finite average separates the two only to the extent that the oscillatory contributions cancel over the given window (figure~\ref{fig:triad-detuning}). The physical meaning of zero frequency, however, depends on the reference frame. In the simple picture of a travelling wave, an observer riding with the wave sees a different slowly varying field from an observer riding with the current and watching the wave pass by. As the comparison in \S\ref{sec:reference-frame-comparison} shows, Eulerian averaging, Lagrangian averaging and the exact projection retain zero-frequency content, but zero frequency has a different physical meaning in each case.

An important outcome of this work is to make explicit that a choice of averaging technique or operator should also be understood as a choice of what can be distinguished over a given time in a given reference frame, because this determines what the resulting wave and mean fields physically represent.

\subsection{The nearly-slow contribution retained by finite averaging}\label{sec:discussion-nearly-slow-drift}

In this work, restricting the infinite-time average to a finite one shows that near-resonant interactions may contribute to the mean when their frequency mismatch is too small to be resolved over that window. To illustrate how these additional near-resonant contributions can affect the dynamics, we consider Stokes drift, computed from correlations between wave-induced displacement and the gradient of the wave velocity \eqref{eq:drift-diagnostic}. In the calculation of \S\ref{sec:drift}, the classical Stokes drift comprises contributions from exactly resonant wave pairs, while the nearly-slow contribution arises from near-resonant frequency pairings. In practice, Stokes drift is often computed by averaging a finite record of data and may therefore include this nearly-slow contribution. This formulation connects Stokes drift, usually treated within Lagrangian averaging, to the resonance-selection mechanism of classical averaging. The two-wave example of \S\ref{sec:andrews-example} isolates this in a particularly simple setting, where adding a second wave at a nearby frequency to a classical single-wave configuration produces a nearly-slow drift that appears as a coherent velocity field.

The broader implication of this example is that the finite window determines whether, and with what weight, a nearly-slow correlation already present in the physical flow enters diagnosed drifts. A finite average can therefore suppress the fast wave oscillations while retaining the nearly-slow correlations that contribute to the drift.

\subsection{Finite-time averaging and near-resonance across physics}\label{sec:discussion-physics-connections}

The view developed here also connects wave--mean decomposition to a broader pattern across physics in which nearby frequencies combine into slowly varying signals and finite observation times set the frequency resolution. A close parallel appears in quantum dynamics, where averaging over a finite interval retains cross terms coupling transitions between a quantum system's energy eigenstates when the differences between their energy gaps cannot be resolved over that interval \citep{schallerbrandes2008}. The same incomplete cancellation occurs in our drift calculation (\S\ref{sec:drift}), where the wave-pair correlations contributing to the nearly-slow drift vary too slowly to cancel appreciably over the averaging window. Finite-time frequency resolution also underlies the derivation of Fermi's golden rule, which gives transition rates between quantum states under a weak perturbation. At finite time, the associated transition-probability profile has the squared sinc form obtained by taking the squared magnitude of the response in \eqref{eq:near-matching-sinc}; as the interval grows, the profile narrows to exact energy conservation \citep[\S~18.2]{shankar1994}. The corresponding infinite-time selection appears in Koopman spectral analysis, which extracts frequencies and modes from nonlinear dynamics by representing the evolution of observables with a linear operator \citep{arbabimezic2017}. Mean-ergodic averaging isolates the invariant, zero-frequency component of those observables \citep{vonneumann1932}. Here, \eqref{eq:app-cesaro-convergence} realizes the same selection directly on the state evolving under the fixed fast generator.

The combination of nearby frequencies into a slowly varying signal also underlies techniques in signal processing, where heterodyne detection provides a lower-frequency readout of a high-frequency signal. For example, mixing the signal with a known reference in a detector whose output is proportional to the square of its input produces this readout at their difference frequency \citep{verbiestrost2015}. In nonlinear optics, difference-frequency generation uses the same cross-term mechanism for frequency conversion, with two incident fields driving a nonlinear medium to produce a new field at their frequency difference \citep[\S~1.2.4]{boyd2020}; in our calculation, the corresponding component appears in the displacement--velocity product \eqref{eq:product-pairing}, which is averaged to obtain the drift.

Our use of the wave reference frame to separate the fast oscillations before averaging (\S\ref{sec:governing-fast-structure}) has a close counterpart in the guiding-centre reduction of charged-particle motion, where rapid gyromotion is separated from the slower drift of the orbit centre \citep{littlejohn1981,caryBrizard2009guidingCentre}. Another connection to plasma physics arises in quasilinear transport, where collisions \citep{duartegorelenkovwhiteberk2019} and dynamical decorrelation \citep{biankontarratcliffe2014} broaden the exact-resonance condition into a finite band within which interactions contribute to transport, analogous to the near-resonant band retained by our finite-window average \eqref{eq:near-matching-general}.

A related relaxation of exact resonance appears in finite wave systems, where quasi-resonant interactions with small frequency mismatch \citep{kartashova2007,kartashova2009} can supplement the exact-resonant network with additional pathways for coherent exchange \citep{kartashova1994}. In the present construction, near-resonant contributions are weighted according to their phase cancellation over the averaging window \eqref{eq:near-matching-general}, rather than selected through a separately prescribed detuning band. Taken together, these connections place near-resonant retention and the slow responses generated by nearby frequencies within a broader pattern recurring across physics.

\subsection{Future directions}\label{sec:discussion-future-directions}

The machinery underlying classical averaging is quite general, as this broader pattern illustrates, but previous work has identified limitations in particular applications. A related difficulty appears in normal-mode initialization, where setting the fast modal amplitudes to zero does not keep them zero under nonlinear evolution, and the initial fast-mode data are corrected order by order to suppress the resulting high-frequency oscillations \citep[\S~3]{baertribbia1977}. Comparisons with GLM-based wave--mean theory have assessed rigorous fast-wave averaging as a route to a closed leading-order balanced evolution over order-one advective times \citep{thomasbuhlershafersmith2018}. In that role, Stokes and pseudomomentum corrections and related wave-induced balance changes are lower order in the strict fast-wave limit and therefore fall outside the leading closure. Although negligible over order-one advective times, those corrections can accumulate over longer times \citep[\S~6]{buhler2000}. Here, instead, we use the same averaging procedure to obtain one such correction directly: the Stokes drift at leading order in wave amplitude for the standard GLM choice of zero-mean particle displacement, with the mean taken here over the fast evolution \citep{vanneste2022stokes}. By averaging the transformed vector field over a finite window, we obtain a closed evolution equation for the full amplitude (\ref{eq:finite-window-closed-evolution}; Appendix~\ref{app:closed-finite-window-dynamics}), retaining near-resonant interactions with evolving phases and weights set by the averaging kernel. Examining the wider mean-flow response associated with these near-resonant interactions would also require accounting for pseudomomentum, whose curl enters the Lagrangian-mean vorticity response in stratified three-dimensional flow \citep{buhlermcintyre1998}.

The analysis presented in this paper assumes that the fast evolution conserves energy and that the wave frequencies remain fixed over each averaging window. The fast motion conserves energy while the wave phases evolve (the skew-adjoint fast operator \(\mathcal L_{\mathrm F}\) generates a unitary group). A non-normal linearization, whose modes need not be orthogonal, can instead amplify disturbances over finite times; its finite-time average is therefore not determined by phase mismatch alone \citep{schmid2007}. The assumption of fixed frequencies excludes passage through resonance, in which the detuning varies with the background and crosses zero. Near such a crossing, the standard averaging construction becomes singular and a separate asymptotic treatment is required \citep[chap.~8]{sandersVerhulstMurdock2007averaging}. Because the local frequency and wavenumber of a slowly modulated wave packet can vary with an inhomogeneous or evolving background \citep{brethertonGarrett1968wavetrains,whitham1974linearNonlinearWaves}, including the mean flow induced by the packet itself \citep{sutherland2006,tabaei2007}, extending the present wave--mean decomposition to such packets would require resolving their slow spatial modulation alongside their fast temporal oscillations. The need to resolve multiple spatial scales also arises in other flows. For instance, the axial dynamics of rapidly rotating convection can vary on both convective and layer-depth scales \citep{calkinsjulienmarti2013}. Extending the present construction to systems in which rapid temporal oscillations coexist with variation across distinct spatial scales is a natural direction for future work.

\subsection{Conclusion}\label{sec:discussion-conclusion}

To conclude, the view of averaging developed here shows that (i) classical, or infinite-time, averaging over the fast-wave evolution produces an exact wave--mean separation of an instantaneous state, in which zero-frequency (non-oscillatory) content is intuitively identified as the non-wave part of the flow, i.e. the mean, but the chosen reference frame determines which dynamics it represents; (ii) when averaging nonlinear wave interactions, restricting the average to a finite window broadens exact resonance selection to a band around zero detuning, retaining interactions whose relative phases change too little to cancel over the time available; and (iii) when calculating Stokes drift, finite-time averaging can retain wave-generated effects that infinite-time averaging removes, including the nearly-slow drift produced by near-resonant wave--wave interactions.

The value of establishing the equivalence between the balanced projection and infinite-time averaging over the fast-wave evolution lies in clarifying the physical balance defining the mean and enabling us to account for the finite duration over which we distinguish it from wave motion. By relating how different averages represent balanced motion, coherent waves and wave-induced transport to the frames in which they are taken and their temporal weighting, it becomes clear how the unified view of averages presented here can inform the choice of mean for capturing phenomena of interest. The fundamental goal of wave--mean decomposition is to separate waves and mean, but any average used to make that separation implicitly depends on a reference frame and therefore embodies a physical meaning of the mean.

\begin{bmhead}[Acknowledgements.]
We thank the Geophysical Fluid Dynamics Summer Program at Woods Hole Oceanographic Institution, where our collaboration began. B.W. would like to thank the University of California, Santa Cruz, for hosting her sabbatical, and the University of Exeter.
\end{bmhead}

\begin{bmhead}[Funding.]
This material is based upon work supported by the Air Force Office of Scientific Research under award number FA9550-23-F-0014 in the amount of \$129,600. This work was also supported by UKRI EPSRC grant EP/R029628/1, Leverhulme Trust Project RPG-2017-098, and Leverhulme Trust Fellowship RF-2022-013.
\end{bmhead}

\begin{bmhead}[Declaration of interests.]
The authors report no conflict of interest.
\end{bmhead}

\begin{bmhead}[Data availability statement.]
Code and supporting data for this study are available on GitHub at \url{https://github.com/lulabels/wave-mean-flow-averaging}, with an archived release at \url{https://doi.org/10.5281/zenodo.23825819}.
\end{bmhead}

\begin{bmhead}[Use of artificial intelligence.]
Large language models were used to generate and modify code for this study, assist in identifying relevant literature, and proofread the manuscript.
\end{bmhead}

\appendix
\renewcommand{\theHsection}{appendix.\Alph{section}}
\section{The exact wave filter as a spectral projection}\label{app:spectral-projection}

Following the periodic, divergence-free energy setting of \citet[\S2.4]{embidmajda1998}, let $\mathcal H$ be the subspace of $L^2(\mathbb T^3;\mathbb C^4)$ consisting of states $\bm w=(\bm u,\rho)^\top$, where $\bm u=(u,v,w)^\top$, with $\nabla\cdot\bm u=0$ in the distributional sense, i.e.\ $\int_{\mathbb T^3}\bm u\cdot\nabla\varphi\,\mathrm{d}\bm x=0$ for every $\varphi\in C^\infty(\mathbb T^3)$. For $\bm w_j=(\bm u_j,\rho_j)^\top$, $j=1,2$, we equip $\mathcal H$ with the energy inner product
\begin{equation}\label{eq:app-energy-inner-product}
\langle \bm w_1,\bm w_2\rangle_{\mathcal H}
\defeq\int_{\mathbb T^3}\bm w_2^\dagger\bm w_1\mathrm{d}\bm x,
\end{equation}
where $\dagger$ denotes conjugate transpose.
The map $\bm u\mapsto\nabla\cdot\bm u$ is continuous from $L^2(\mathbb T^3;\mathbb C^3)$ to $H^{-1}(\mathbb T^3)$, so its kernel is closed. Hence $\mathcal H$ is a closed subspace of $L^2(\mathbb T^3;\mathbb C^4)$ and is therefore a Hilbert space. Let $\mathbb P:L^2(\mathbb T^3;\mathbb C^3)\to L^2(\mathbb T^3;\mathbb C^3)$ denote the $L^2$-orthogonal Leray projection onto $\ker(\nabla\cdot)$, given by
\begin{equation}\label{eq:app-Leray-projection}
\mathbb P\bm a\defeq\bm a-\nabla\Delta^{-1}\nabla\cdot\bm a.
\end{equation}
Here $\Delta^{-1}$ is the inverse Laplacian, understood with the usual zero-mean periodic convention.
The fast operator in \eqref{eq:LF-qg-em98} can then be written as
\begin{equation}\label{eq:app-LF-Leray-form}
\mathcal L_{\mathrm F}\bm w
\defeq\left(\mathbb P(F\hat{\bm e}_z\times\bm u+\rho\hat{\bm e}_z),-w\right)^\top.
\end{equation}
The projection $\mathbb P$ supplies the pressure correction required to preserve incompressibility. Indeed,
\begin{subequations}\label{eq:app-Leray-verification}
\begin{align}
\nabla\cdot\left(F\hat{\bm e}_z\times\bm u+\rho\hat{\bm e}_z\right)
&=-F\hat{\bm e}_z\cdot(\nabla\times\bm u)+\partial_z\rho,
\label{eq:app-Leray-divergence}\\
\mathbb P\left(F\hat{\bm e}_z\times\bm u+\rho\hat{\bm e}_z\right)
&=F\hat{\bm e}_z\times\bm u+\rho\hat{\bm e}_z+\nabla\Delta^{-1}\left(F\hat{\bm e}_z\cdot(\nabla\times\bm u)-\partial_z\rho\right).
\label{eq:app-Leray-recovery}
\end{align}
\end{subequations}
Equation~\eqref{eq:app-Leray-recovery} is precisely the velocity component in \eqref{eq:LF-qg-em98}; together with the unchanged density component $-w$, it verifies that \eqref{eq:app-LF-Leray-form} reproduces the full operator $\mathcal L_{\mathrm F}\bm w$.
Since $\mathbb P$ is an $L^2$-orthogonal projection, $\|\mathbb P\bm a\|_{L^2}\leq\|\bm a\|_{L^2}$ for every $\bm a\in L^2(\mathbb T^3;\mathbb C^3)$. Applying this inequality to \eqref{eq:app-LF-Leray-form} gives
\begin{align}
\|\mathcal L_{\mathrm F}\bm w\|_{\mathcal H}^2
&=\left\|\mathbb P\left(F\hat{\bm e}_z\times\bm u+\rho\hat{\bm e}_z\right)\right\|_{L^2}^2+\|w\|_{L^2}^2\nonumber\\
&\leq\left\|F\hat{\bm e}_z\times\bm u+\rho\hat{\bm e}_z\right\|_{L^2}^2+\|w\|_{L^2}^2\nonumber\\
&=F^2\|\hat{\bm e}_z\times\bm u\|_{L^2}^2+\|\rho\|_{L^2}^2+\|w\|_{L^2}^2\nonumber\\
&\leq(F^2+1)\|\bm u\|_{L^2}^2+\|\rho\|_{L^2}^2\nonumber\\
&\leq(F^2+1)\|\bm w\|_{\mathcal H}^2.
\label{eq:app-LF-bound}
\end{align}
Indeed, $F\hat{\bm e}_z\times\bm u=(-Fv,Fu,0)^\top$ and $\rho\hat{\bm e}_z=(0,0,\rho)^\top$, so they are pointwise orthogonal. Moreover, $|\hat{\bm e}_z\times\bm u|^2=|u|^2+|v|^2\leq|\bm u|^2$ and $|w|^2\leq|\bm u|^2$, which yield the fourth line after integration. The final inequality follows from $F^2+1\geq1$ and, by \eqref{eq:app-energy-inner-product}, $\|\bm w\|_{\mathcal H}^2=\|\bm u\|_{L^2}^2+\|\rho\|_{L^2}^2$. Thus $\mathcal L_{\mathrm F}$ is bounded on $\mathcal H$. Since $\mathbb P$ is self-adjoint and acts as the identity on divergence-free velocities, while the Coriolis map and the map $(w,\rho)^\top\mapsto(\rho,-w)^\top$ arising from the terms $\rho\hat{\bm e}_z$ and $-w$ in \eqref{eq:app-LF-Leray-form} are skew-Hermitian, $\mathcal L_{\mathrm F}$ satisfies
\begin{equation}\label{eq:app-LF-skew-symmetry}
\langle\mathcal L_{\mathrm F}\bm w_1,\bm w_2\rangle_{\mathcal H}
=-\langle \bm w_1,\mathcal L_{\mathrm F}\bm w_2\rangle_{\mathcal H},
\qquad \bm w_1,\bm w_2\in\mathcal H.
\end{equation}
Since $\mathcal L_{\mathrm F}$ is bounded on $\mathcal H$, the Riesz representation theorem ensures that its adjoint is defined on all of $\mathcal H$ and is uniquely characterized by $\langle\mathcal L_{\mathrm F}\bm w_1,\bm w_2\rangle_{\mathcal H}=\langle\bm w_1,\mathcal L_{\mathrm F}^*\bm w_2\rangle_{\mathcal H}$. Comparing this identity with \eqref{eq:app-LF-skew-symmetry} gives $\mathcal L_{\mathrm F}^*\bm w=-\mathcal L_{\mathrm F}\bm w$ for every $\bm w\in\mathcal H$, and hence $\mathcal L_{\mathrm F}^*=-\mathcal L_{\mathrm F}$. Consequently, $-i\mathcal L_{\mathrm F}$ is self-adjoint and, by Stone's theorem, $U(t)\defeq e^{t\mathcal L_{\mathrm F}}$ is a strongly continuous one-parameter unitary group on $\mathcal H$.

Because $-i\mathcal L_{\mathrm F}$ is self-adjoint, the spectral theorem provides a unique projection-valued measure $E_{\mathrm F}$ on $\mathbb R$ such that, for every bounded Borel function $f:\mathbb R\to\mathbb C$,
\begin{equation}\label{eq:app-borel-spectral}
f(-i\mathcal L_{\mathrm F}) = \int_{\mathbb{R}} f(\omega)\mathrm{d}E_{\mathrm F}(\omega).
\end{equation}
For a Borel set $B\subset\mathbb R$, the operator $E_{\mathrm F}(B)$ projects onto the fast-frequency spectral subspace corresponding to $B$.
At zero frequency, the spectral subspace is $\ker(-i\mathcal L_{\mathrm F})=\ker\mathcal L_{\mathrm F}$. We therefore define its orthogonal projection by
\begin{equation}\label{eq:app-P-def}
P \defeq E_{\mathrm F}(\{0\}).
\end{equation}
Since the spectral projection $P$ commutes with $U(t)$ and $U(t)$ acts as the identity on $\operatorname{ran}P$ (the fast evolution leaves the zero-frequency subspace fixed), we have $U(t)P=PU(t)=P$ for every $t\in\mathbb R$.

To prove \eqref{eq:exact-wave-filter}, consider the finite-time orbit-average operator $(1/T)\int_0^T e^{-s\mathcal L_{\mathrm F}}\mathrm{d}s$ on its right-hand side. Its spectral multiplier is
\begin{equation}\label{eq:app-cesaro-multiplier}
f_T(\omega) \defeq \frac{1}{T}\int_0^T e^{-is\omega}\mathrm{d}s.
\end{equation}
By \eqref{eq:app-borel-spectral}, the corresponding orbit-average operator has the spectral representation
\begin{equation}\label{eq:app-cesaro-spectral-representation}
\frac{1}{T}\int_0^T e^{-s\mathcal L_{\mathrm F}}\mathrm{d}s
=f_T(-i\mathcal L_{\mathrm F})
=\int_{\mathbb R}f_T(\omega)\mathrm{d}E_{\mathrm F}(\omega).
\end{equation}
For $\omega=0$, $f_T(0)=1$. For $\omega\neq0$, evaluating the integral in \eqref{eq:app-cesaro-multiplier} gives
\begin{equation}\label{eq:app-cesaro-nonzero-evaluation}
f_T(\omega)
=\frac{1-e^{-iT\omega}}{iT\omega}.
\end{equation}
Since $|1-e^{-iT\omega}|\leq2$, \eqref{eq:app-cesaro-nonzero-evaluation} gives (for $\omega\neq0$)
\begin{equation}\label{eq:app-cesaro-nonzero-limit}
\lim_{T\to\infty}|f_T(\omega)|
=\lim_{T\to\infty}\frac{|1-e^{-iT\omega}|}{T|\omega|}
=0.
\end{equation}
Therefore, for every $\omega\in\mathbb R$,
\begin{equation}\label{eq:app-cesaro-pointwise-limit}
\lim_{T\to\infty}f_T(\omega)
=\begin{cases}
1, & \omega=0,\\
0, & \omega\neq0,
\end{cases}
=\mathbf 1_{\{0\}}(\omega).
\end{equation}
For an arbitrary $\bm w\in\mathcal{H}$, let $\mu_{\bm w}(B)\defeq\|E_{\mathrm F}(B)\bm w\|_{\mathcal H}^2$. This finite scalar measure records the energy of $\bm w$ in the frequency set $B$, with $\mu_{\bm w}(\mathbb R)=\|\bm w\|_{\mathcal H}^2$. From \eqref{eq:app-borel-spectral}, $\|g(-i\mathcal L_{\mathrm F})\bm w\|_{\mathcal H}^2=\int_{\mathbb R}|g(\omega)|^2\mathrm{d}\mu_{\bm w}(\omega)$ for every bounded Borel function $g$. Because $|f_T(\omega)|\leq 1$, the pointwise limit \eqref{eq:app-cesaro-pointwise-limit} and the dominated convergence theorem give
\begin{equation}\label{eq:app-cesaro-convergence}
\left\|\left[f_T(-i\mathcal L_{\mathrm F})-P\right]\bm w\right\|_{\mathcal H}^2=\int_{\mathbb R}\left|f_T(\omega)-\mathbf 1_{\{0\}}(\omega)\right|^2\mathrm{d}\mu_{\bm w}(\omega)\xrightarrow[T\to\infty]{}0.
\end{equation}
Since \eqref{eq:app-cesaro-convergence} holds for every $\bm w\in\mathcal H$, the orbit-average operators converge strongly to $P$; this is precisely \eqref{eq:exact-wave-filter}. Thus, as $T\to\infty$, the orbit average converges to the projection onto the $U(s)$-invariant subspace $\ker\mathcal L_{\mathrm F}$, while every nonzero fast-frequency spectral component vanishes.

In the QG scaling \eqref{eq:qg-operators-em98}, the nonzero fast-wave frequencies satisfy $|\omega|\geq\delta\defeq\min\{1,F\}>0$. Since the averaging error vanishes at zero frequency and \eqref{eq:app-cesaro-nonzero-evaluation} bounds its multiplier by $2/(\delta T)$ elsewhere on the spectrum, the spectral representation \eqref{eq:app-cesaro-convergence} gives
\begin{equation}\label{eq:app-cesaro-operator-norm-bound}
\left\|\frac{1}{T}\int_0^T e^{-s\mathcal L_{\mathrm F}}\mathrm{d}s-P\right\|_{\mathcal H\to\mathcal H}
\leq\frac{2}{\delta T}.
\end{equation}
Thus, for fixed $F>0$, convergence to $P$ holds in operator norm at rate $O(T^{-1})$.

Having established the exact infinite-time projection in \eqref{eq:exact-wave-filter}, we now recover the finite-window multiplier in \eqref{eq:spectral-multiplier}. Let $K\in L^1(\mathbb R)$ be normalized as in \S\ref{sec:exact-and-finite}. Since the fast evolution is unitary, the kernel-weighted orbit is integrable, and applying the spectral representation \eqref{eq:app-borel-spectral} under the integral gives
\begin{equation}\label{eq:app-kernel-average}
\bigl(A_T(e^{-\tau\mathcal L_{\mathrm F}}\bm w)\bigr)(\tau_0)
=\int_{\mathbb R}K_T(s)e^{-(\tau_0+s)\mathcal L_{\mathrm F}}\bm w\mathrm{d}s
=e^{-\tau_0\mathcal L_{\mathrm F}}\int_{\mathbb R}\widehat K(T\omega)\mathrm{d}E_{\mathrm F}(\omega)\bm w.
\end{equation}
Thus each fast-frequency component at frequency $\omega$ is weighted by $\widehat K(T\omega)$, which is precisely \eqref{eq:spectral-multiplier}.

\section{Averaging and generator identities}\label{app:averaging-generator-identities}

\subsection{The sinc response of a centered boxcar average}\label{app:boxcar-sinc-response}

Changing variables from the time offset $s$ to the absolute time $t=t_0+s$ in \eqref{eq:centered-uniform-average} gives
\begin{equation}\label{eq:app-boxcar-def}
(A_T f)(t_0)
= \frac{1}{T}\int_{t_0-T/2}^{t_0+T/2} f(t)\mathrm{d}t.
\end{equation}
Applying \eqref{eq:app-boxcar-def} to a pure oscillation $f(t)=e^{-i\omega t}$, for $\omega\neq0$, gives
\begin{align}
(A_T f)(t_0)
&= \frac{1}{T}\int_{t_0-T/2}^{t_0+T/2} e^{-i\omega t}\mathrm{d}t \notag\\
&= \frac{1}{T}\left[\frac{e^{-i\omega t}}{-i\omega}\right]_{t_0-T/2}^{t_0+T/2} \notag\\
&= \frac{e^{-i\omega t_0}}{T}\frac{e^{-i\omega T/2}-e^{i\omega T/2}}{-i\omega} \notag\\
&= e^{-i\omega t_0}\frac{\sin(\omega T/2)}{\omega T/2}
= e^{-i\omega t_0}\operatorname{sinc}\left(\frac{\omega T}{2}\right).
\end{align}
At $\omega=0$, the integrand is constant, so both sides equal one.

\subsection{The material derivative as the generator of trajectory evolution}

To derive \eqref{eq:material-characteristic-evolution}, let $\bm{X}(s;\bm{x},t)$ denote the material trajectory satisfying
\begin{equation}\label{eq:app-material-trajectory}
\frac{\mathrm d}{\mathrm ds}\bm{X}(s;\bm{x},t)=\bm{u}\bigl(\bm{X}(s;\bm{x},t),t+s\bigr),\qquad \bm{X}(0;\bm{x},t)=\bm{x}.
\end{equation}
The chain rule gives
\begin{align}
\frac{\mathrm d}{\mathrm ds}\theta\bigl(\bm{X}(s;\bm{x},t),t+s\bigr)
&=\bigl(\partial_t\theta+\bm{u}\cdot\nabla\theta\bigr)\bigl(\bm{X}(s;\bm{x},t),t+s\bigr)\notag\\
&=\bigl(D_{\mathrm L}\theta\bigr)\bigl(\bm{X}(s;\bm{x},t),t+s\bigr).\label{eq:app-material-chain-rule}
\end{align}
Equation~\eqref{eq:app-material-chain-rule} shows that $D_{\mathrm L}$ generates the evolution of scalar fields along the material trajectory. Just as $y'=Ay$, with $y(0)=y_0$, has the solution $y(s)=e^{sA}y_0$, the corresponding scalar-field evolution is represented by $e^{sD_{\mathrm L}}$. Using $\bm{X}(0;\bm{x},t)=\bm{x}$, we obtain
\begin{equation}
\bigl(e^{sD_{\mathrm L}}\theta\bigr)(\bm{x},t)=\theta\bigl(\bm{X}(s;\bm{x},t),t+s\bigr),
\end{equation}
as stated in \eqref{eq:material-characteristic-evolution}.

\subsection{Frequency selection by Lagrangian and Eulerian averages}\label{app:frame-frequency-selection}

To obtain the frequency weights  used in \S\ref{sec:reference-frame-comparison}, we Fourier analyse the Lagrangian and Eulerian time histories with respect to the same time-offset variable $s$. With $(\bm{x},t)$ fixed, define
\begin{equation}\label{eq:app-frame-time-dependences}
f_{\mathrm L}(s)\defeq\theta\bigl(\bm{X}(s;\bm{x},t),t+s\bigr),\qquad f_{\mathrm E}(s)\defeq\theta(\bm{x},t+s).
\end{equation}
These are different functions of the same offset $s$: $f_{\mathrm L}$ samples the field along a material trajectory, whereas $f_{\mathrm E}$ samples it at the fixed position $\bm{x}$.
Because the finite-window average is linear, it suffices, for $R\in\{\mathrm L,\mathrm E\}$, to evaluate it on a single Fourier component $f_R(s)=a_R e^{-i\omega_Rs}$. Here $a_R\in\mathbb C$ is independent of $s$, and any dependence on the fixed base point $(\bm{x},t)$ is suppressed. Applying the finite-window average gives
\begin{align}
(A_T^R\theta)(\bm{x},t)&=\int_{\mathbb{R}}K_T(s)f_R(s)\mathrm{d}s\notag\\
&=a_R\int_{\mathbb{R}}K_T(s)e^{-i\omega_Rs}\mathrm{d}s\notag\\
&=a_R\widehat{K_T}(\omega_R).\label{eq:app-frame-frequency-response}
\end{align}
Hence each Lagrangian or Eulerian frequency is weighted by $\widehat{K_T}(\omega_R)$; by linearity, this weighting applies mode by mode to any Fourier superposition.

\section{Closed dynamics under finite-window averaging}\label{app:closed-finite-window-dynamics}

To derive \eqref{eq:finite-window-closed-evolution}, apply the transformation into the wave reference frame, $\tilde{\bm{w}}\defeq e^{\tau\mathcal{L}_{\mathrm F}}\bm{w}$, with $\tau\defeq t/\varepsilon$, to \eqref{eq:abstract-system}. Differentiation gives
\begin{equation}\label{eq:app-wave-frame-cancellation}
\begin{aligned}
\partial_t\tilde{\bm{w}}
&=\frac{1}{\varepsilon}e^{\tau\mathcal{L}_{\mathrm F}}\mathcal{L}_{\mathrm F}\bm{w}
+e^{\tau\mathcal{L}_{\mathrm F}}\Bigl(-\frac{1}{\varepsilon}\mathcal{L}_{\mathrm F}\bm{w}-\mathcal{L}_S\bm{w}-\mathcal{B}(\bm{w},\bm{w})+\mathcal{D}(\bm{w})\Bigr)\\
&=-e^{\tau\mathcal{L}_{\mathrm F}}\Bigl(\mathcal{L}_S(e^{-\tau\mathcal{L}_{\mathrm F}}\tilde{\bm{w}})+\mathcal{B}(e^{-\tau\mathcal{L}_{\mathrm F}}\tilde{\bm{w}},e^{-\tau\mathcal{L}_{\mathrm F}}\tilde{\bm{w}})-\mathcal{D}(e^{-\tau\mathcal{L}_{\mathrm F}}\tilde{\bm{w}})\Bigr).
\end{aligned}
\end{equation}
The singular fast terms have cancelled exactly. Thus the averaged vector field in \eqref{eq:averaged-equation} is the infinite fast-orbit average of the slow linear, nonlinear, and diffusive terms evaluated on $e^{-s\mathcal{L}_{\mathrm F}}\overline{\bm{w}}$ and mapped back by $e^{s\mathcal{L}_{\mathrm F}}$. Since the transformed vector field in \eqref{eq:app-wave-frame-cancellation} is evaluated at the current fast phase $\tau$, we form its finite-window average by replacing $\tau$ with $\tau+s$ and integrating over $s$, while holding the amplitude fixed. Using the normalized kernel $K_T(s)\defeq (1/T)K(s/T)$, where $K\in L^1(\mathbb{R})$ and $\int_{\mathbb{R}}K(s)\mathrm{d}s=1$, gives the finite-window evolution \eqref{eq:finite-window-closed-evolution}:
\begin{equation}\label{eq:app-closed-finite-window-evolution}
\partial_t\overline{\bm{w}}_{K,T}+\int_{\mathbb{R}}K_T(s)e^{(\tau+s)\mathcal{L}_{\mathrm F}}\Bigl((\mathcal{L}_S-\mathcal{D})e^{-(\tau+s)\mathcal{L}_{\mathrm F}}\overline{\bm{w}}_{K,T}+\mathcal{B}(e^{-(\tau+s)\mathcal{L}_{\mathrm F}}\overline{\bm{w}}_{K,T},e^{-(\tau+s)\mathcal{L}_{\mathrm F}}\overline{\bm{w}}_{K,T})\Bigr)\mathrm{d}s=0.
\end{equation}
The amplitude $\overline{\bm{w}}_{K,T}$ is defined as the solution of \eqref{eq:app-closed-finite-window-evolution}. The finite-window evolution is closed on the full amplitude $\overline{\bm{w}}_{K,T}$, but the evolution of $P\overline{\bm{w}}_{K,T}$ generally depends on the retained fast component $(I-P)\overline{\bm{w}}_{K,T}$.

On the periodic domain considered here, we expand $\overline{\bm{w}}_{K,T}$ in Fourier eigenmodes of $\mathcal{L}_{\mathrm F}$, with $\beta$ labeling the different eigenmodes at each wavevector:
\begin{equation}\label{eq:app-finite-window-mode-expansion}
\overline{\bm{w}}_{K,T}(\bm{x},t)=\sum_{\bm{k},\beta}\sigma_{\bm{k}}^{\beta}(t)e^{i\bm{k}\cdot\bm{x}}r_{\bm{k}}^{\beta},\qquad \mathcal{L}_{\mathrm F}(\bm{k})r_{\bm{k}}^{\beta}=i\omega_{\bm{k}}^{\beta}r_{\bm{k}}^{\beta}.
\end{equation}
For a quadratic interaction in which modes $(\bm{k}_1,\beta_1)$ and $(\bm{k}_2,\beta_2)$ generate mode $(\bm{k}_3,\beta_3)$, with $\bm{k}_1+\bm{k}_2=\bm{k}_3$, the detuning is
\begin{equation}\label{eq:app-finite-window-detuning}
\Delta\omega\defeq \omega_{\bm{k}_1}^{\beta_1}+\omega_{\bm{k}_2}^{\beta_2}-\omega_{\bm{k}_3}^{\beta_3}.
\end{equation}
Substituting \eqref{eq:app-finite-window-mode-expansion} into the quadratic term of \eqref{eq:app-closed-finite-window-evolution} gives
\begin{equation}\label{eq:app-conjugated-quadratic-modal}
\begin{aligned}
&e^{(\tau+s)\mathcal{L}_{\mathrm F}}\mathcal{B}\left(e^{-(\tau+s)\mathcal{L}_{\mathrm F}}\overline{\bm{w}}_{K,T},e^{-(\tau+s)\mathcal{L}_{\mathrm F}}\overline{\bm{w}}_{K,T}\right)\\
&=\sum_{\bm{k}_3}\sum_{\bm{k}_1+\bm{k}_2=\bm{k}_3}\sum_{\beta_3,\beta_1,\beta_2}B_{\bm{k}_3,\bm{k}_1,\bm{k}_2}^{\beta_3,\beta_1,\beta_2}\sigma_{\bm{k}_1}^{\beta_1}(t)\sigma_{\bm{k}_2}^{\beta_2}(t)e^{-i(\tau+s)\Delta\omega}e^{i\bm{k}_3\cdot\bm{x}}r_{\bm{k}_3}^{\beta_3},
\end{aligned}
\end{equation}
where $B_{\bm{k}_3,\bm{k}_1,\bm{k}_2}^{\beta_3,\beta_1,\beta_2}$ is the corresponding modal interaction coefficient. Writing the phase in \eqref{eq:app-conjugated-quadratic-modal} as $e^{-i(\tau+s)\Delta\omega}=e^{-i\tau\Delta\omega}e^{-is\Delta\omega}$, the factor $e^{-i\tau\Delta\omega}$ is independent of $s$ and can be taken outside the integral. Integrating the remaining phase against $K_T$ in \eqref{eq:app-closed-finite-window-evolution} gives
\begin{equation}\label{eq:app-finite-window-triad-weight}
\begin{aligned}
&\int_{\mathbb{R}}K_T(s)e^{(\tau+s)\mathcal{L}_{\mathrm F}}\mathcal{B}\Bigl(e^{-(\tau+s)\mathcal{L}_{\mathrm F}}\overline{\bm{w}}_{K,T},e^{-(\tau+s)\mathcal{L}_{\mathrm F}}\overline{\bm{w}}_{K,T}\Bigr)\mathrm{d}s\\
&=\sum_{\bm{k}_3}\sum_{\bm{k}_1+\bm{k}_2=\bm{k}_3}\sum_{\beta_3,\beta_1,\beta_2}B_{\bm{k}_3,\bm{k}_1,\bm{k}_2}^{\beta_3,\beta_1,\beta_2}\sigma_{\bm{k}_1}^{\beta_1}(t)\sigma_{\bm{k}_2}^{\beta_2}(t)e^{-i\tau\Delta\omega}\widehat{K}(T\Delta\omega)e^{i\bm{k}_3\cdot\bm{x}}r_{\bm{k}_3}^{\beta_3}.
\end{aligned}
\end{equation}
Hence, integration against $K_T$ replaces the varying factor $e^{-is\Delta\omega}$ by the detuning weight $\widehat{K}(T\Delta\omega)$ used in \S\ref{sec:exact-and-finite}. The remaining factors are independent of $s$ and remain unchanged, including $e^{-i\tau\Delta\omega}$, which describes the interaction's phase at the current time. The exact averaged equation instead weights each quadratic interaction by $\mathbf{1}_{\{\Delta\omega=0\}}$.
The product $T\Delta\omega$ is the phase mismatch accumulated over the averaging window. If $T|\Delta\omega|=O(1)$, the weight $\widehat K(T\Delta\omega)$ need not decay; if $T|\Delta\omega|\ll1$, it remains close to $\widehat K(0)=1$. Thus interactions with $|\Delta\omega|\ll T^{-1}$ are retained with nearly unit weight; at $|\Delta\omega|=O(T^{-1})$, their weight depends on $K$. The corresponding state-frequency multiplier is $\widehat K(T\omega)$ (Appendix~\ref{app:spectral-projection}). Since $K\in L^1(\mathbb R)$, the Riemann--Lebesgue lemma gives $\widehat K(\xi)\to0$ as $|\xi|\to\infty$. Hence, $\widehat K(T\Delta\omega)\to0$ for each fixed $\Delta\omega\neq0$ as $T\to\infty$.

\end{document}